\documentclass[preprint]{aastex631}

\newcommand{\mod}[1]{\textbf{}}

\newcommand{\ratioeq}{\langle B_{t}^{2} \rangle/\langle B_{0}^{2} \rangle}
\newcommand{\ratioinline}{\frac{\langle B_{t}^{2} \rangle}{\langle B_{0}^{2} \rangle}}
\newcommand{\dispfunct}{1 - \langle\cos[\Delta \phi(\ell)] \rangle}
\newcommand{\kms}{km~s$^{-1}$}
\shorttitle{DCF and Shear Flows}
\shortauthors{Guerra et al.}
\graphicspath{{./}{}}

\begin{document}

\title{The Strength of the Sheared Magnetic Field in the Galactic's Circum-Nuclear Disk}

\author[0000-0001-8819-9648]{Jordan A. Guerra}
\affil{Department of Physics, Villanova University, 800 E. Lancaster Ave., Villanova, PA 19085, USA}

\author[0000-0001-5357-6538]{Enrique Lopez-Rodriguez}
\affil{Kavli Institute for Particle Astrophysics \& Cosmology (KIPAC), Stanford University, Stanford, CA 94305, USA}

\author[0000-0003-0016-0533]{David T. Chuss}
\affil{Department of Physics, Villanova University, 800 E. Lancaster Ave., Villanova, PA 19085, USA}

\author[0000-0002-4013-6469]{Natalie O. Butterfield}
\affiliation{National Radio Astronomy Observatory, 520 Edgemont Road, Charlottesville, VA 22903, USA}

\author[0000-0001-7908-6940]{Joan T. Schmelz}
\affil{USRA, 425 3rd Street SW, Suite 950, Washington, DC, 20024, USA}

\correspondingauthor{Jordan A. Guerra}
\email{jordan.guerraaguilera@villanova.edu}



\begin{abstract}
Recent high-resolution 53-$\mu$m polarimetric observations from SOFIA/HAWC+ have revealed the inferred plane-of-the-sky magnetic field (B-field) orientation in the Galactic center's Circum-Nuclear Disk (CND). The B-field is mostly aligned with the steamers of ionized material falling onto Sgr A* at large, differential velocities (shear). In such conditions, estimating the B-field strength with the ``classical" Davis-Chandrasekhar-Fermi (DCF) method does not provide accurate results. We derive a ``modified'' DCF method by solving the ideal MHD equations from first principles considering the effects of a large-scale, shear flow on the propagation of a fast magnetosonic wave. In the context of the DCF approximation, both the value of the shear and its Laplacian affect the inferred B-field strength. Using synthetic polarization data from MHD simulations for a medium dominated by shear flows, we find that the ``classical'' DCF determines B-field strengths only within $>50$\% of the true value where the ``modified" DCF results are improved significantly ($\sim$3-22\%). Applying our ``modified'' DCF method to the CND revealed B-field strengths of 1 - 16 mG in the northern arm, 1 - 13 mG in the eastern arm, and 3 - 27 mG in the western arm at spatial scales $\lesssim1$ pc, with median values of $5.1\pm0.8$, $4.0\pm1.2$, and $8.5\pm2.3$ mG, respectively. The balance between turbulent gas energy (kinetic plus hydrostatic) and turbulent magnetic energy densities suggest that, along the magnetic-field-flow direction, magnetic effects become less dominant as the shear flow increases and weakens the B-field via magnetic convection. Our results indicate that the transition from magnetically to gravitationally dominated accretion of material onto Sgr A* starts at distances $\sim$ 1 pc.\end{abstract}



\section{Introduction} \label{sec:intro}

The Galactic Center, with its supermassive blackhole, many-faceted structures, multi-waveband emission, strong magnetic fields, and high-velocity gas, may be (arguably) the most exotic environment in the Milky Way. The Circum-Nuclear Disk (CND) \citep[e.g.,][]{Becklin1982,Liszt1983,Guesten1987,McGary2001} is an inclined, $\sim70^{\circ}$, ring of dense, $10^{4}-10^{6}$ cm$^{-3}$, molecular gas and dust with inner and outer radii of $\sim1.5$ pc and $3-7$ pc, respectively  \citep[e.g.,][]{Guesten1987,Jackson1993,Etxaluze2011,Lau2013}. This structure is dynamically complex, with molecular and ionized gas streamers feeding the blackhole, which is coincident with the radio source designated Sgr~A$^{*}$ \citep{Hsieh2017,Hsieh2021}. The molecular gas within the central few pc is influenced by the gravitational potential of Sgr~A$^{*}$ and follows an almost Keplerian rotation. 

Despite this extreme complexity, a spiral magnetic field (B-field) morphology has been reported using the 850-$\mu$m polarimetry of the James Clerk Maxwell Telescope \citep[JCMT;][]{Hsieh2018}. At $\sim15\arcsec$ ($0.38$ pc) resolution, the $850~\mu$m polarimetric observations trace the B-field associated with magnetically aligned dust grains in the cold and dense interstellar medium (ISM) cospatial with the molecular gas streamers feeding Sgr~A$^{*}$. The CND has been observed using $53~\mu$m polarimetric observations \citep{Dowell2019} with the High-resolution Airborne Wideband Camera Plus \citep[HAWC+,][]{Harper2018} instrument on the 2.7-m Stratospheric Observatory for Infrared Astronomy  \citep[SOFIA,][]{Temi2018}. At an angular resolution of $4.85\arcsec$ ($0.18$ pc), 53-$\mu$m polarization data imply that the inferred B-field orientation follows a set of three spiral structures. These preliminary results suggest that matter may be flowing along B-field lines or that the B-field is being sheared by differential motions of matter. The B-fields may be an important contributor to the removal of the angular momentum for the gas flowing onto Sgr~A$^{*}$ and are expected to be in close equipartition with the turbulent kinetic and cosmic-ray energies. Thus, an energy budget across the multi-faceted structures of the CND is required to put the magnetohydrodynamical (MHD) inflow onto Sgr~A$^{*}$ into context.

The plane-of-the-sky (POS) B-field strength can be estimated using the Davis-Chandrasekhar-Fermi (DCF) method \citep{Davis1951,Chandrasekhar1953}, which relates the line-of-sight (LOS) velocity dispersion and the POS polarization angle dispersion. It assumes an isotropically turbulent medium whose turbulent kinetic and turbulent magnetic energy components are (approximately) in equipartition. The original or $``$classical$"$ DCF method assumes a steady state with no large-scale flows. The small angular dispersion of the polarization angle orientations of the HAWC+ data resulted in a preliminary estimate of the B-field strength of $\sim$5 milliGauss (mG) throughout the $\sim5$ pc CND  \citep{Dowell2019}. This value is similar to Zeeman results for the LOS component of 2 mG from \citet{Killeen1992} and 0.6 - 3 mG from \citet{Plante1995}.

\citet{Schmelz2020} used these results to estimate the plasma beta ($\beta$), the ratio of the thermal-to-magnetic pressure. This value is used traditionally as an indicator of whether magnetic or thermodynamic processes dominate in an environment. If the thermal pressure is greater than the magnetic pressure, $\beta>1$, referred to as a high-$\beta$ plasma, the gas dynamics will control the structure of the environment, e.g., the solar photosphere. If the thermal pressure is less than the magnetic pressure, $\beta<1$, referred to as a low-$\beta$ plasma, the B-field will control the structure of the environment, e.g., the solar corona. Thus, this parameter can help distinguish between the two possible scenarios for the CND. Using the preliminary value of the B-field strength indicates that $\beta \sim$ 0.001. However, the widths of all molecular, atomic, and ionized gas lines are quite large in and around the CND, indicating that the turbulence-associated energy density can be larger than the thermal energy density. For example, using up to 30 rotational H$_{2}$ lines observed with the Infrared Space Observatory, \citet{Mills2017} found velocity dispersions, $\sigma_{v}$, from 31 – 112 km s$^{-1}$ for regions in the CND. Thus a more appropriate ratio is $\beta^\prime$ defined as the ratio of the turbulent kinetic pressure over the magnetic pressure. Using an average of the values calculated by \citet{Mills2017}, $\sigma_{v}$ = 44.4 km s$^{-1}$, it was estimated that $\beta^\prime \sim$ 0.03 -- an order of magnitude larger than $\beta$. These values are clearly in the low-$\beta$ regime where the magnetic pressure dominates. Similar results were obtained by \citet{Hsieh2018} with $\beta\sim0.7$, assuming a B-field strength of $1$ mG, a thermal pressure of $P_{\rm th}/k\sim 4\times10^{8}$ cm$^{-3}$ K, an electron density of $7.8\times10^{4}$ cm$^{-3}$, and an electron temperature of $5000$ K. These results indicate that, like the solar corona, the B-field is channeling the plasma and appears to be a significant force on the matter in this region. 

However, these preliminary results are limited by the assumptions inherent in the classical DCF approach, i.e., an ideal medium in a steady state with no large-scale flows and incompressible B-fields. Several authors have proposed modifications to the DCF method \citep{Hildebrand2009,Houde2009}, but these only improve the estimation of the angle dispersion. Using an energy balance, \citet{Skalidis2021} have recently derived a modified DCF method for compressible modes of the B-fields valid for no self-gravity-dominated environments. This modification has been numerically supported by \citet{Beattie2022}. Furthermore, \citet{Lopez2021} investigated how the DCF estimation of the B-field could be affected by the strong outflow from the core of the starburst galaxy, Messier 82. From first principles, they solved the ideal-MHD equations for a medium with a combination of turbulent B-fields and a large-scale B-field in the same direction as the steady flow. They determined that the large-scale flow can increase or decrease the B-field strength from the classical DCF depending if the isotropic turbulent velocity dispersion is larger or smaller than the flow itself. Whether the B-fields in the CND are affected by a large-scale flow and/or are compressed and sheared by the gas streamers is still unclear. These configurations need to be characterized and applied to a modify version of the original DCF method to obtain scientifically meaningful B-field strengths.

In this paper, we take another step away from the ideal conditions of the original DCF. We develop a wave equation and dispersion relation for the case where an Alfv\'en wave is propagating in the presence of structured (position-dependent) steady state flows. This allows us to investigate the effects of shearing by differential motions of matter on the B-field in the inner 5 pc of the Milky Way. Section \ref{sec:wave_eq} presents the derivation of the wave equation and dispersion relation. Section \ref{sec:apps} shows how the wave dispersion relation can be used in the DCF context to estimate the B-field strength. The modified DCF expression is first tested using synthetic data (Section \ref{sec:synth_data}) and then used to estimate the POS B-field strength in the CND (Section \ref{sec:CND}). We discuss our results in Section \ref{sec:discussion} and our conclusions are summarized in Section \ref{sec:summary}. 

\section{Modified DCF Method} \label{sec:wave_eq}

The DCF method is a well-known approximation to estimate the strength of the POS magnetic field from the dispersion patterns observed in the dust polarization directions. This approximation results from assuming that an Alfv\'en wave with dispersion relation

\begin{equation}
    \omega^{2} = v_{A}^{2}k^{2}
    \label{eq:disp_rel_ideal}
\end{equation}

\noindent
exist in the magnetized gas. Eq. \ref{eq:disp_rel_ideal} correspond to a wave propagating in an ideal medium with velocity equal to the global Alfv\'en velocity $v_{A}=B/\sqrt{4\pi\rho}$, where $B$ is the strength of the magnetic field and $\rho$ is the mass density. Assuming that the turbulence is isotropic, $\omega$ can be related to the LOS velocity dispersion $\sigma_{v}$ (due to turbulent motions) and the wave vector $k$ to the deviation of the polarization angles due to turbulence, $\sigma_{\phi}$. Thus Eq. \ref{eq:disp_rel_ideal} becomes

\begin{equation}
    \sigma_{v}^{2} = v_{A}^{2}\sigma_{\phi}^{2}.
    \label{eq:disp_rel_ideal_1}
\end{equation}

Using the definition of Alfv\'en speed, we can obtain from Eq. \ref{eq:disp_rel_ideal_1} the well-known DCF expression

\begin{equation}
    B_{\rm POS}^{\rm DCF} = \sqrt{4\pi\rho}\frac{\sigma_{v}}{\sigma_{\phi}}.
    \label{eq:DCF_og}
\end{equation}

However, Eq. \ref{eq:DCF_og} (or Eq.\ref{eq:disp_rel_ideal}) does not consider cases where the Alfv\'en wave might be propagating in a non-ideal medium such as a plasma with significant viscosity or in the presence of a steady-state flow. In the following we develop a wave equation and dispersion relation that considers the latter. Throughout this paper, we shall refer to Eq. \ref{eq:DCF_og} as the $``$classical$"$ DCF.

\subsection{Governing Equations} \label{subsec:tables}

We start from the magnetohydrodynamics (MHD) equations of a cold (plasma $\beta \rightarrow 0$), non-dissipative medium

\begin{equation}
    \frac{\partial\rho}{\partial t} + \nabla\cdot(\rho\mathbf{v})=0, \quad \text{(Continuity)}
    \label{eq:cont_eq}
\end{equation}

\begin{equation}
    \rho\left[\frac{\partial \mathbf{v}}{\partial t} + (\mathbf{v}\cdot\nabla)\mathbf{v}\right] = -\frac{1}{4\pi}\mathbf{B}\times(\nabla\times\mathbf{B}), \quad \text{(Momentum)}
    \label{eq:momentum_eq}
\end{equation}

\begin{equation}
    \nabla\cdot\mathbf{B} = 0, \quad \text{(Divergence - Free)}
    \label{eq:div_eq}
\end{equation}

\begin{equation}
    \frac{\partial\mathbf{B}}{\partial t} = \nabla\times(\mathbf{v}\times\mathbf{B}). \quad \text{(Induction)}
    \label{eq:diffusion_eq}
\end{equation}

\noindent
where $\rho$ is the mass density, $\mathbf{v}$ is the velocity, and $\mathbf{B}$ is the magnetic field. In order to linearise these equations, we choose to express them in Cartesian coordinates, where ($x,y$) correspond to the POS plane and the $z$ coordinate is perpendicular, or in the LOS. However, in this work we will not consider the evolution of variables in the $z$ direction, so the $z$-component of vector quantities are set to zero and any derivative in this direction vanishes ($\partial_{z}=0$).

Assuming that each MHD variable can be decomposed into a time-independent background (steady state) plus a time-dependent perturbation

\begin{equation}
    \begin{array}{c}
        \rho \rightarrow \rho_{0} + \rho,  \\
         \mathbf{v} \rightarrow \mathbf{U}_{0} + \mathbf{v}, \\
         \mathbf{B} \rightarrow \mathbf{B}_{0} + \mathbf{B}.
    \end{array}
    \label{eq:perturbations}
\end{equation}

We further assume that the steady state ($\rho_{0},\mathbf{U}_{0},\mathbf{B}_{0}$) corresponds to constant background density, a constant magnetic field in the $\hat{x}$-direction, $\mathbf{B}_{0} = B_{0}\hat{x}$, and a parallel position-dependent background flow, $\mathbf{U}_{0} = U_{0}(x,y)\hat{x}$.

Introducing the expression in Eq. \ref{eq:perturbations} into Eqs. \ref{eq:cont_eq} - \ref{eq:diffusion_eq} and keeping only terms up to first order in perturbations, we obtain

\begin{equation}
    \frac{\partial\rho}{\partial t}+\frac{\partial}{\partial x}(\rho_{0}U_{0}) +\frac{\partial}{\partial x}(\rho_{0}v_{x}) +\frac{\partial}{\partial y}(\rho_{0}v_{y}) +\frac{\partial}{\partial x}(\rho U_{0})= 0, \quad \text{(Continuity)}
    \label{eq:lin_cont}
\end{equation}

\begin{equation}
    \rho_{0}\left[ \frac{\partial v_{x}}{\partial t} +U_{0}\frac{\partial U_{0}}{\partial x} +\left( v_{x}\frac{\partial U_{0}}{\partial x} + v_{y}\frac{\partial U_{0}}{\partial y}\right) +U_{0}\frac{\partial v_{x}}{\partial x}\right] +\rho U_{0}\frac{\partial U_{0}}{\partial x} = 0, \quad \text{(Momentum~x)}
    \label{eq:lin_mom_x}
\end{equation}

\begin{equation}
    \rho_{0}\left( \frac{\partial v_{y}}{\partial t} + U_{0}\frac{\partial v_{y}}{\partial {\bf x}}\right) = \frac{B_{0}}{4\pi}\left( \frac{\partial B_{y}}{\partial x} - \frac{\partial B_{x}}{\partial y}\right), \quad \text{(Momentum~y)}
    \label{eq:lin_mom_y}
\end{equation}

\begin{equation}
    \frac{\partial B_{0}}{\partial x} + \frac{\partial B_{x}}{\partial x} + \frac{\partial B_{y}}{\partial y} = 0,  \quad \text{(Divergence~free)}
    \label{eq:lin_div0}
\end{equation}

\begin{equation}
    \frac{\partial B_{x}}{\partial t} = \frac{\partial}{\partial y}(U_{0}B_{y}-v_{y}B_{0}), \quad \text{(Induction~x)}
    \label{eq:lin_diff_x}
\end{equation}

\begin{equation}
    \frac{\partial B_{y}}{\partial t} = {\bf -}\frac{\partial}{\partial x}(U_{0}B_{y}-v_{y}B_{0}). \quad \text{(Induction~y)}
    \label{eq:lin_diff_y}
\end{equation}

Eqs. \ref{eq:lin_cont} - \ref{eq:lin_diff_y} represent the system of a coupled linear fast and slow magnetosonic waves.

\subsection{The Wave Equation}

In order to study the propagation of the fast wave, we combine Eqs. \ref{eq:lin_mom_y}, \ref{eq:lin_diff_x}, \ref{eq:lin_diff_y} in order to find a wave equation only for the $y$ component of each variable. First, we take the temporal derivative of Eq. \ref{eq:lin_mom_y}. Then, assuming the variables are smooth and well-behaved, we can interchange temporal and spatial derivatives. Finally, substituting Eqs. \ref{eq:lin_diff_x} and \ref{eq:lin_diff_y} in the right hand side, resulting in

\begin{equation}
    \frac{\partial^{2}v_{y}}{\partial t^{2}} + U_{0}\frac{\partial}{\partial t}\left(\frac{\partial v_{y}}{\partial x} \right) = v_{A}^{2}\left\{ \frac{\partial^{2}v_{y}}{\partial x^{2}} + \frac{\partial^{2}v_{y}}{\partial y^{2}} - \frac{1}{B_{0}}\left[\frac{\partial^{2}(U_{0}B_{y})}{\partial x^{2}} + \frac{\partial^{2}(U_{0}B_{y})}{\partial y^{2}}\right] \right\}.
    \label{eq:wave_eq}
\end{equation}

Eq.~\ref{eq:wave_eq} corresponds to the fast magnetosonic wave propagating in the ($x,y$) plane with at the global Alfv\'en speed. The dispersion relation from which a modified DCF expression can be derived from, is found by assuming the perturbed variables as plane wave which propagate entirely in the $x-$direction with a $k_{x}-$vector that is affected by the structure of the steady-state flow. That is,

\begin{equation}
    \begin{array}{c}
         v_{y} = v_{y,0}\exp[i(k_{x}(x,y)x - \omega t)],  \\
         B_{y} = B_{y,0}\exp[i(k_{x}(x,y)x - \omega t)],
    \end{array}
\end{equation}

\noindent
into Eq.~\ref{eq:wave_eq}. The resulting dispersion relation is

\begin{equation}
    \label{eq:disp_rel}
    -\omega^{2} + \omega U_{0}\left( k_{x} + x\frac{\partial k}{\partial x}\right) = v_{A}^{2}\mathcal{F}(k_{x}),
\end{equation}

\noindent
where $\mathcal{F}$ is a complex function of $k_{x}$ and $U_{0}$ with real and imaginary parts
\begin{equation}
    -\Re[\mathcal{F}(k_{x})] = \left(1 {\bf -} \frac{B_{y,0}}{B_{0}}\frac{U_{0}}{v_{y,0}}\right)\left\{k_{x}^{2} + 2k_{x}x\frac{\partial k_{x}}{\partial x} + x^{2}\left[ \left( \frac{\partial k_{x}}{\partial x}\right)^{2} + \left( \frac{\partial k_{x}}{\partial y}\right)^{2}\right]\right\} {\bf +} \frac{B_{y,0}}{B_{0}v_{y,0}} \left( \frac{\partial^2 U_{0}}{\partial x^{2}} + \frac{\partial^2 U_{0}}{\partial y^{2}}\right),
    \label{eq:F_real}
\end{equation}

\begin{equation}
    \Im[\mathcal{F}(k_{x})] = \left(1 {\bf -} \frac{B_{y,0}}{B_{0}}\frac{U_{0}}{v_{y,0}}\right)\left[ 2\frac{\partial k_{x}}{\partial x} + x\left( \frac{\partial^{2} k_{x}}{\partial x^{2}} + \frac{\partial^{2} k_{x}}{\partial y^{2}}\right)\right] {\bf -} 2\frac{B_{y,0}}{B_{0}v_{y,0}} \left[ \frac{\partial U_{0}}{\partial x}\left( k_{x} + x\frac{\partial k_{x}}{\partial x}\right) {\bf +} x\frac{\partial U_{0}}{\partial y}\frac{\partial k_{x}}{\partial y}\right],
    \label{eq:F_img}
\end{equation}

Since the left-hand-side of Eq. \ref{eq:disp_rel} is a real function, thus it follows that

\begin{equation}
    -\omega^{2} + \omega U_{0}\left(k_{x}+x\frac{\partial k_{x}}{\partial x}\right) = v_{A}^{2}\Re[\mathcal{F}(k_{x})],
    \label{eq:disp_rel_real}
\end{equation}

\begin{equation}
    0 = v_{A}^{2}\Im[\mathcal{F}(k_{x})].
    \label{eq:disp_rel_img}
\end{equation}

Thus, Eq.~\ref{eq:disp_rel_real} is the wave dispersion relation from which we can derive the modified DCF approximation. Eq. \ref{eq:disp_rel_img}, on the other hand, might provide some constrains on the values of $k_{x}$ and their relationship to $U_{0}$. Re-writing Eq. \ref{eq:disp_rel_real} as

\begin{equation}
    \label{eq:new_va}
    v_{A}^{2} = \frac{\omega^{2} - \omega U_{0}\left(k_{x}+x\frac{\partial k_{x}}{\partial x}\right)}{-\Re[\mathcal{F}(k)]},
\end{equation}

\noindent
we see that the Alfv\'en speed --  and therefore the magnetic field strength -- depends explicitly on the distance along the direction of the steady-state flow ($x$-direction) and implicitly on the perpendicular coordinate ($y$) through the spatial dependence of the wave vector and the steady flow.

\subsection{Applications} \label{sec:apps}

Following the classical DCF method \citep{Davis1951,Chandrasekhar1953}, we can make the substitutions $\omega \rightarrow \sigma_{v}$, $k_{x} \rightarrow \sigma_{\phi}$ in Eq. \ref{eq:new_va}. Similarly, we can make the substitutions $v_{y,0} \rightarrow \sigma_{v}$ and $B_{y,0}/B_{0} \rightarrow \sqrt{\ratioinline} = \sigma_{\phi}$ \citep{Houde2009,Hildebrand2009}. Along with the definition of $v_{A}$, the position-dependent strength of a magnetic field being affected by steady-state flow is

\begin{equation}
    B_{\rm POS} = \sqrt{4\pi\rho}\sigma_{v}\left|\frac{1 - (U_{0}/\sigma_{v})(\sigma_{\phi}+x\frac{\partial\sigma_{\phi}}{\partial x})}{\left[ 1 - \sigma_{\phi}\frac{U_{0}}{\sigma_{v}}\right]\left\{ \sigma_{\phi}^{2} + 2x\sigma_{\phi}\frac{\partial \sigma_{\phi}}{\partial x} + x^{2}\left[ \left( \frac{\partial\sigma_{\phi}}{\partial x}\right)^{2} + \left( \frac{\partial\sigma_{\phi}}{\partial y}\right)^{2}\right]\right\} + \frac{\sigma_{\phi}}{\sigma_{v}}\nabla^{2}U_{0}}\right|^{1/2}.
    \label{eq:dcf_shf_long}
\end{equation}

\noindent
where we have taken the absolute value since the Alfv\'en velocity is a real number (Eq. \ref{eq:new_va}). In order to use Eq.~\ref{eq:dcf_shf_long} with polarimetric data, it is necessary to know the variation of $\sigma_{\phi}$ with coordinates $x$ and $y$. In contrast, if $\sigma_{\phi}$ corresponds to an average value (calculated over the region where $U_{0}$ and $B_{0}$ are parallel to each other), then its derivatives vanish. This simplification results in the expression for the shear-flow modified DCF expression

\begin{equation}
    B_{\rm POS}^{\rm DCF, SF} = \sqrt{4\pi\rho}\sigma_{v} \left| \frac{1 - \frac{U_{0}}{\sigma_{v}}\sigma_{\phi} }{ \sigma_{\phi}^{2} - \sigma_{\phi}^{3}\frac{U_{0}}{\sigma_{v}} + \frac{\sigma_{\phi}}{\sigma_{v}}\nabla^{2}U_{0}}\right|^{1/2}.
    \label{eq:mod_dcf}
\end{equation}

For Eq.~\ref{eq:mod_dcf} to have units of magnetic field, the term inside the root square must be ``dimensionless" which implies that the Laplacian of $U_{0}$ must have units of speed, e.g., using normalized spatial dimensions. It is easy to see that Eq. \ref{eq:mod_dcf} reduces to the classical DCF expression (Eq. \ref{eq:DCF_og}) for $U_{0}=0$, {\it i.e.,} $B_{\rm POS}^{\rm DCF, SF} (U_{0} = 0) = B_{\rm POS}^{\rm DCF}$. As in the large-scale flow correction to the classical DCF expression made by \citet{Lopez2021}

\begin{equation}
        B_{\rm POS}^{\rm DCF, F} = B_{\rm POS}^{\rm DCF}\left| 1 - \sigma_{\phi}\frac{U_{0}}{\sigma_{v}} \right|.
    \label{eq:DCF_lsflow}
\end{equation}

\noindent
Eq.~\ref{eq:mod_dcf} also includes the ratio $U_{0}/\sigma_{v}$ as weight to the terms related to the shear flow. 

The shear-flow modified DCF expression ($B_{\rm POS}^{\rm DCF, SF}$, Eq.~\ref{eq:mod_dcf}) has a more complex functional dependence with $\sigma_{\phi}$ than its classical and large-scale flow counterparts.  Inside the square root, terms that arise because of the presence of $U_{0}$ are proportional to $\sigma^{3}_{\phi}$ and $\sigma_{\phi}$. These terms are weighted by the strength of the shear flow and its Laplacian relative to the strength of the turbulent motions, respectively. By definition $U_{0}(x,y)>0$ (i.e., flow moves in the $+x$-direction), and results in lower values of POS magnetic field. On the other hand, $\nabla^{2}U_{0}$ can be positive (speed increases) or negative (speed decreases). For a fixed value of $\sigma_{\phi}$, positive values of $\nabla^{2}U_{0}$ implies that the effect of $U_{0}(x,y)$ is to decrease the value of $B^{\rm DCF,SF}_{\rm POS}$ while for negative values of $\nabla^{2}U_{0}$, $B^{\rm DCF,SF}_{\rm POS}$ will increase.

\begin{figure}[!h]
    \centering
    \includegraphics[width=0.45\textwidth]{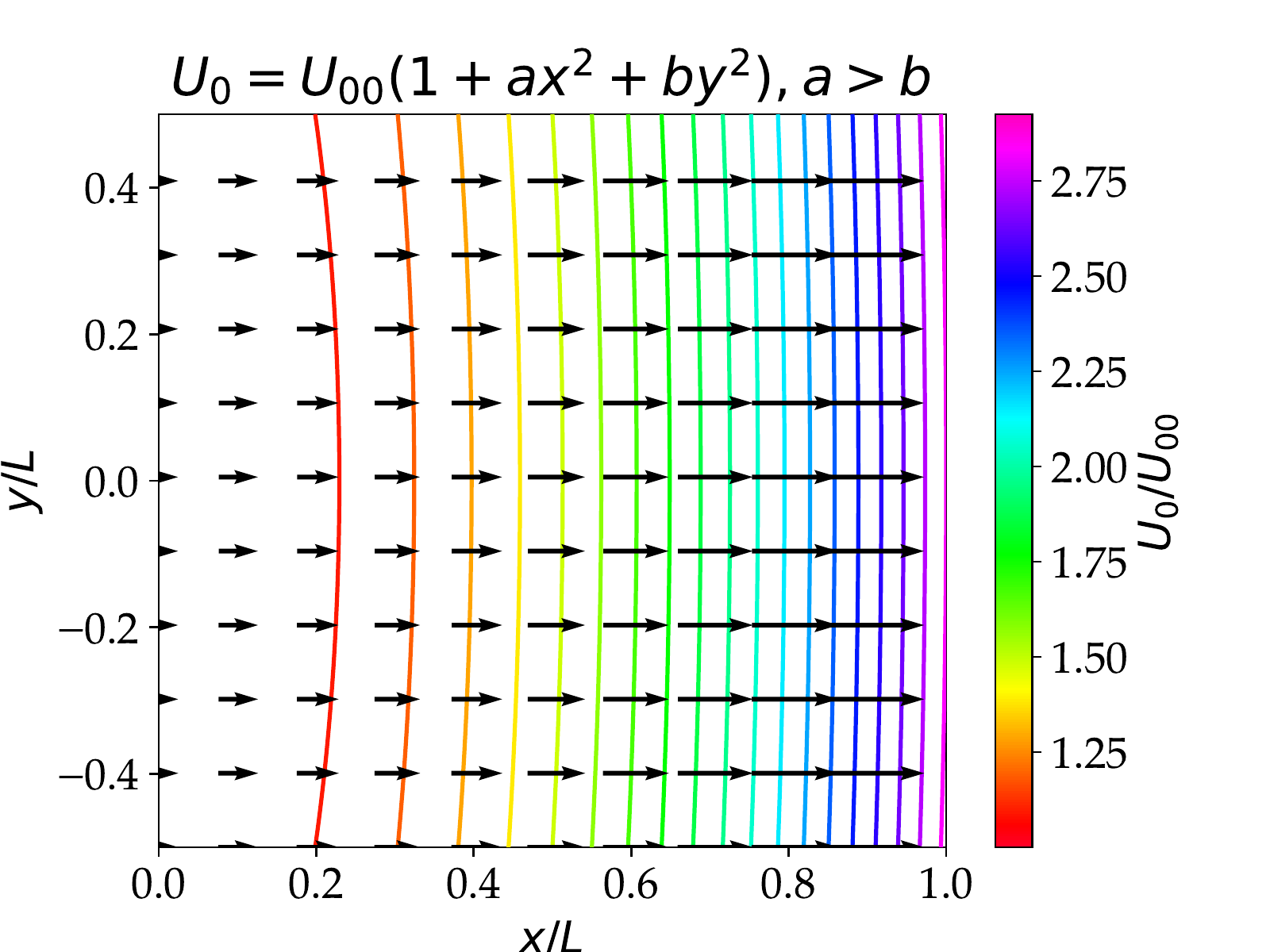}
    \includegraphics[width=0.45\textwidth]{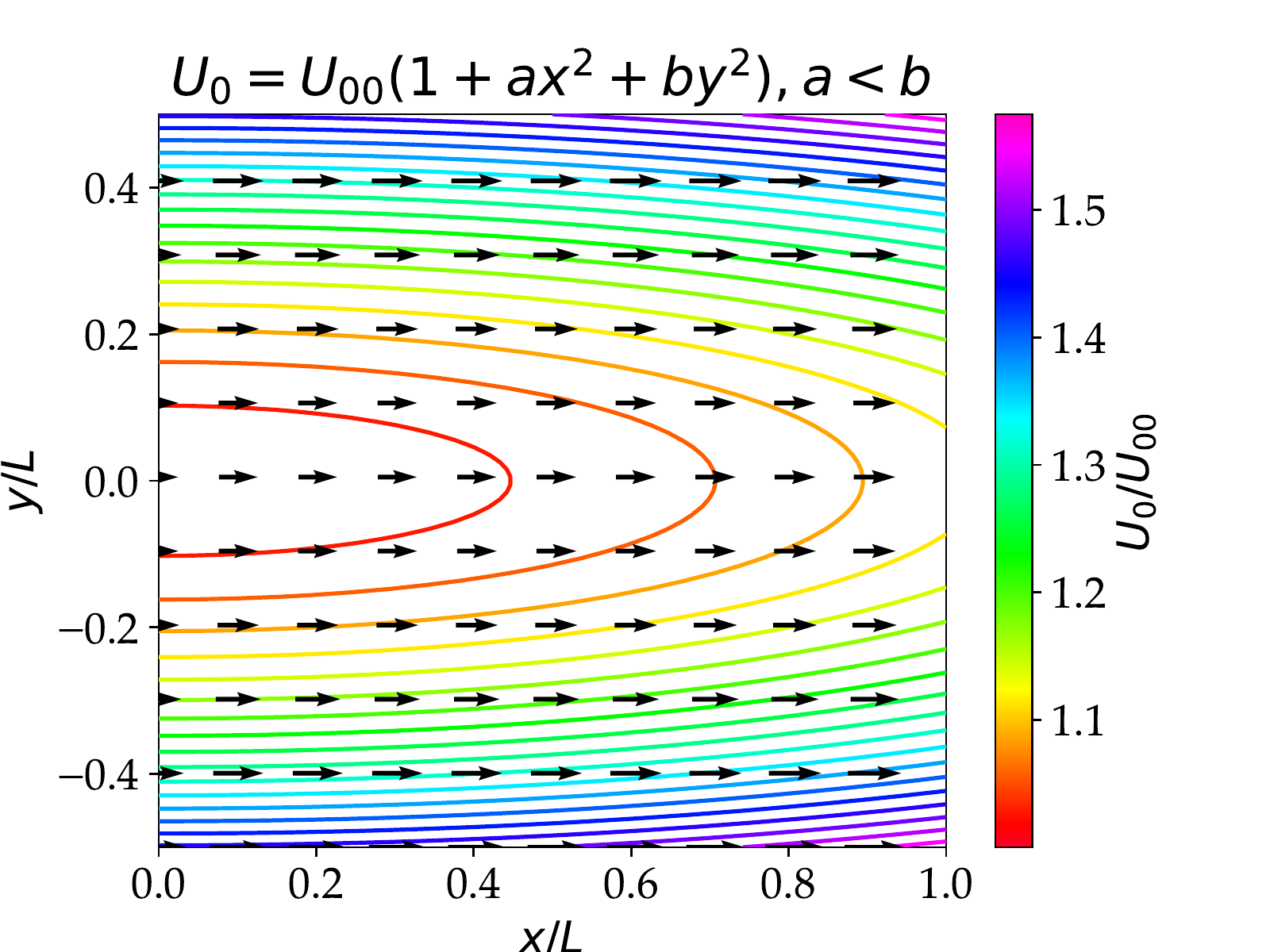}\\
    \includegraphics[width=0.45\textwidth]{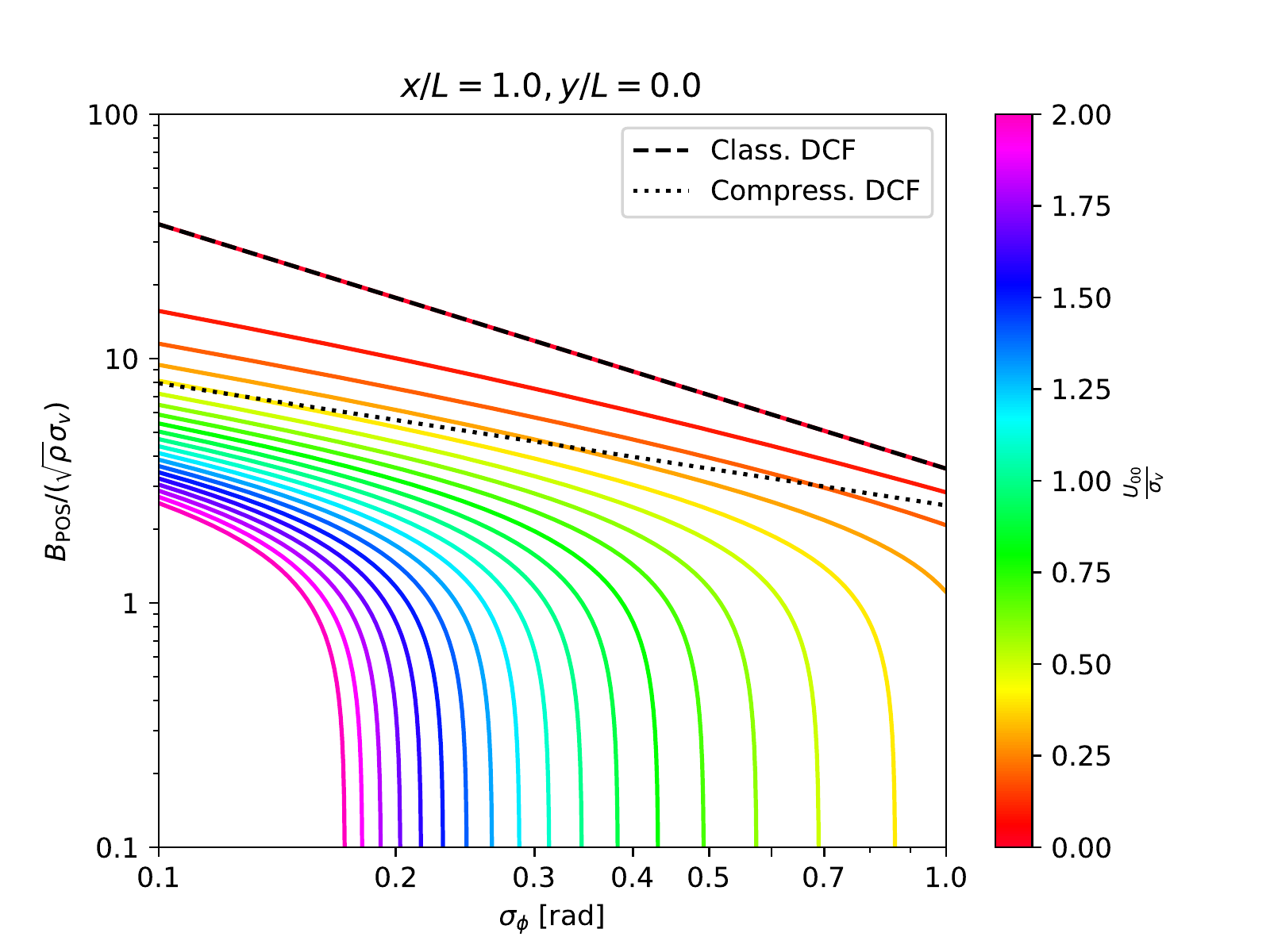}
    \includegraphics[width=0.45\textwidth]{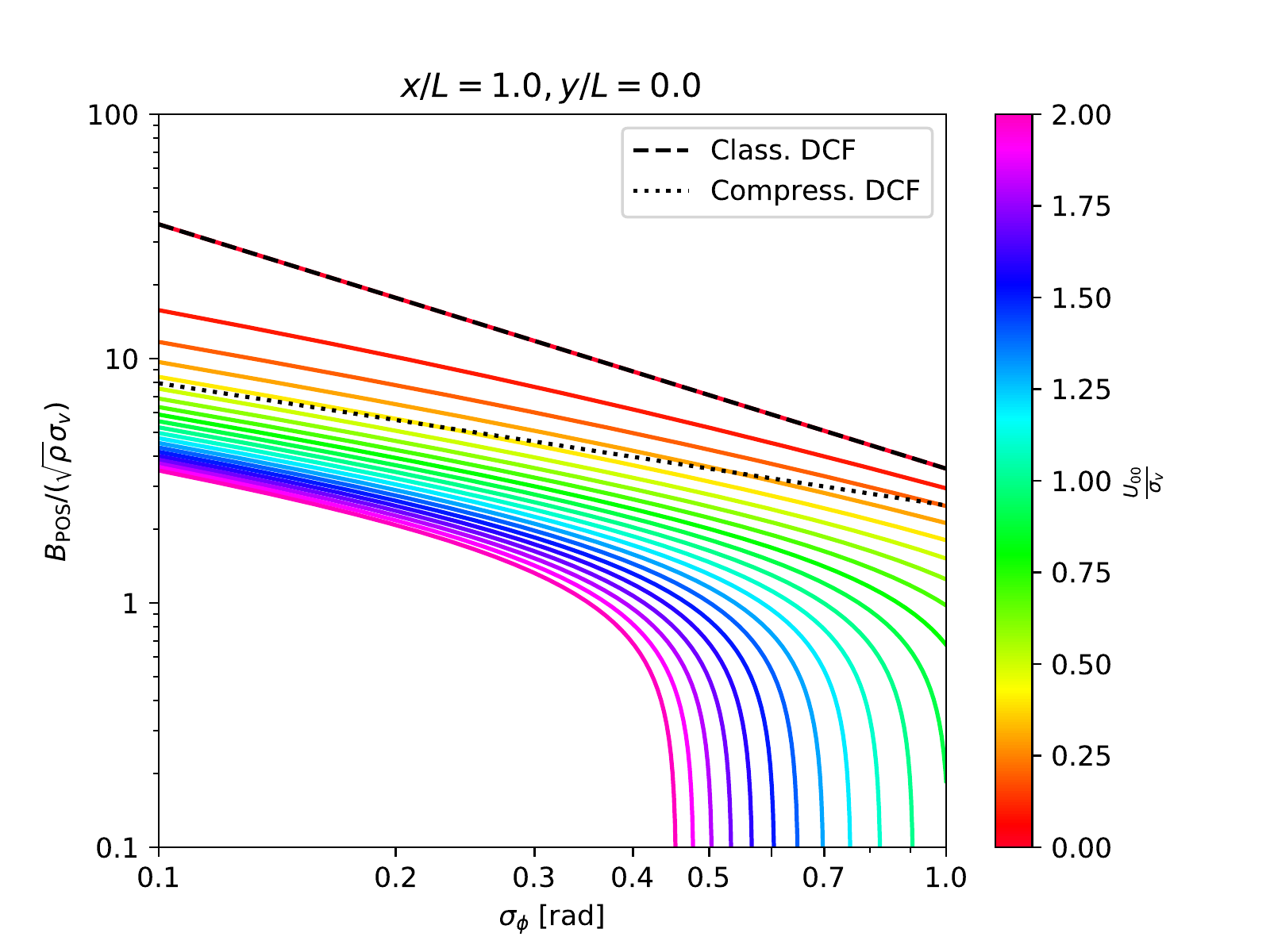}\\
    
    \caption{{\it Top panels:} Steady-state flow, $U_{0}(x,y)$ for two cases: stretch ({\it Left}) and shear ({\it Right}). {\it Bottom panels:} POS magnetic field strength, $B_{\rm POS}$, as a function of the polarization angle dispersion, $\sigma_{\phi}$. The results from the modified DCF approximation for different values of $U_{00}/\sigma_{v}$ -- the ratio of the strength of $U_{0}$ to the velocity dispersion -- are shown in color. For comparison, the classical DCF and the compressional \citep{Skalidis2021} DCF are shown in dashed and dotted black lines, respectively. Values of $B_{\rm POS}$ are shown for $y/L=0$, $x/L=1$, where $L$ is the characteristic length of the flow in the $x$-direction.}
    \label{fig:plots_Bpos}
\end{figure}

Figure \ref{fig:plots_Bpos} displays values of $B_{\rm POS}/\sqrt{\rho}\sigma_{v}$ as a function of $\sigma_{\phi}$ ({\it lower panels}) for two cases of steady-state flow ({\it top panels}). The flow is assumed to be parabolic, $U_{0}(x,y)=U_{00}(1 + ax^{2} + by^{2})$, with strength $U_{00}$, in order to ensure that $\nabla^{2}U_{0}\neq0$ and positive. In Figure \ref{fig:plots_Bpos}, {\it left panels} correspond to $a>b$ which we refer to as {\it stretch} flow (larger increase in $x$), while {\it right panels} correspond to $a<b$ or {\it shear} flow (larger increase in $y$). The spatial coordinates in Figure \ref{fig:plots_Bpos} are normalized by the characteristic length of the flow in the $x$-direction, $L$. In the {\it Bottom panels} several curves are shown: color-scaled curves correspond to different values of the fraction $U_{00}/\sigma_{v}$; dashed black curve displays the values of $B_{\rm POS}$ for the classical DCF approximation (coincidental with that of $U_{00}/\sigma_{v}=0$) and the black dotted curve correspond to the compressional DCF \citep{Skalidis2021}. These panels correspond to values of $x/L=1$ and $y/L=0$. 

We see in the {\it Bottom panels} of Figure \ref{fig:plots_Bpos} that $B_{\rm POS}$ consistently decreases with values $\sigma_{\phi}$. Classical and compressional DCF values, show a steady decrease for all values of angular dispersion. For the modified DCF expression, the value $B_{\rm POS}$ appears to decrease steadily at first -- similarly to the classical or compressional way -- then the value decreases drastically in a asymptotic manner. Such asymptotic value appears at smaller $\sigma_{\phi}$ with the increase of $U_{00}/\sigma_{v}$. This behavior is consistent with the conditions stated above. When comparing the {\it stretch} and {\it shear} cases, we found the scenario just mentioned is true in both cases. However, for the same value $U_{00}/\sigma_{v}$ of, in the shear flow case the asymptote appears at larger values of $\sigma_{\phi}$ compared to the stretch flow. Similar results were found for values of $x/L=0$ but with a weaker dependence on the type of flow, indicating that an spatially-dependent steady-state flow becomes more important (in the context of the DCF approximation) further along direction of the flow. 

More importantly, bottom plots in Figure \ref{fig:plots_Bpos} show that for both type of flows, the magnetic field strength will be overestimated in the presence of a large scale, sheared flow if its velocity field it's not considered according to Eq. \ref{eq:mod_dcf}. As a reminder to the read, Table \ref{tab:table0} shows a summary of the definitions of the several DCF methods derived and applied in this work.

\begin{deluxetable*}{llccc}[ht!]
\tablecaption{Derived definitions of the DCF method for several physical environments.\label{tab:table0}}
\tablewidth{0pt}
\tablehead{\colhead{DCF Method}	&	\colhead{Definition}		& Eq. \\ 
	&	&	&
}
\startdata
Classical	&	$B_{\rm POS}^{\rm DCF} = \sqrt{4\pi\rho}\frac{\sigma_{v}}{\sigma_{\phi}}$ & 	 \ref{eq:DCF_og}\\
Large-scale flow & $B_{\rm POS}^{\rm DCF, F} =  B_{\rm POS}^{\rm DCF}\left| 1 - \sigma_{\phi}\frac{U_{0}}{\sigma_{v}} \right|$ &
\ref{eq:DCF_lsflow} \\
Shear-flow & $B_{\rm POS}^{\rm DCF, SF} = \sqrt{4\pi\rho}\sigma_{v} \left| \frac{1 - \frac{U_{0}}{\sigma_{v}}\sigma_{\phi} }{ \sigma_{\phi}^{2} - \sigma_{\phi}^{3}\frac{U_{0}}{\sigma_{v}} + \frac{\sigma_{\phi}}{\sigma_{v}}\nabla^{2}U_{0}}\right|^{1/2}$ & \ref{eq:mod_dcf}
\enddata
\end{deluxetable*}

\section{Testing the modified DCF method with Synthetic Data}\label{sec:synth_data}

\begin{figure}[!h]
    \centering
    \includegraphics[width=0.45\textwidth]{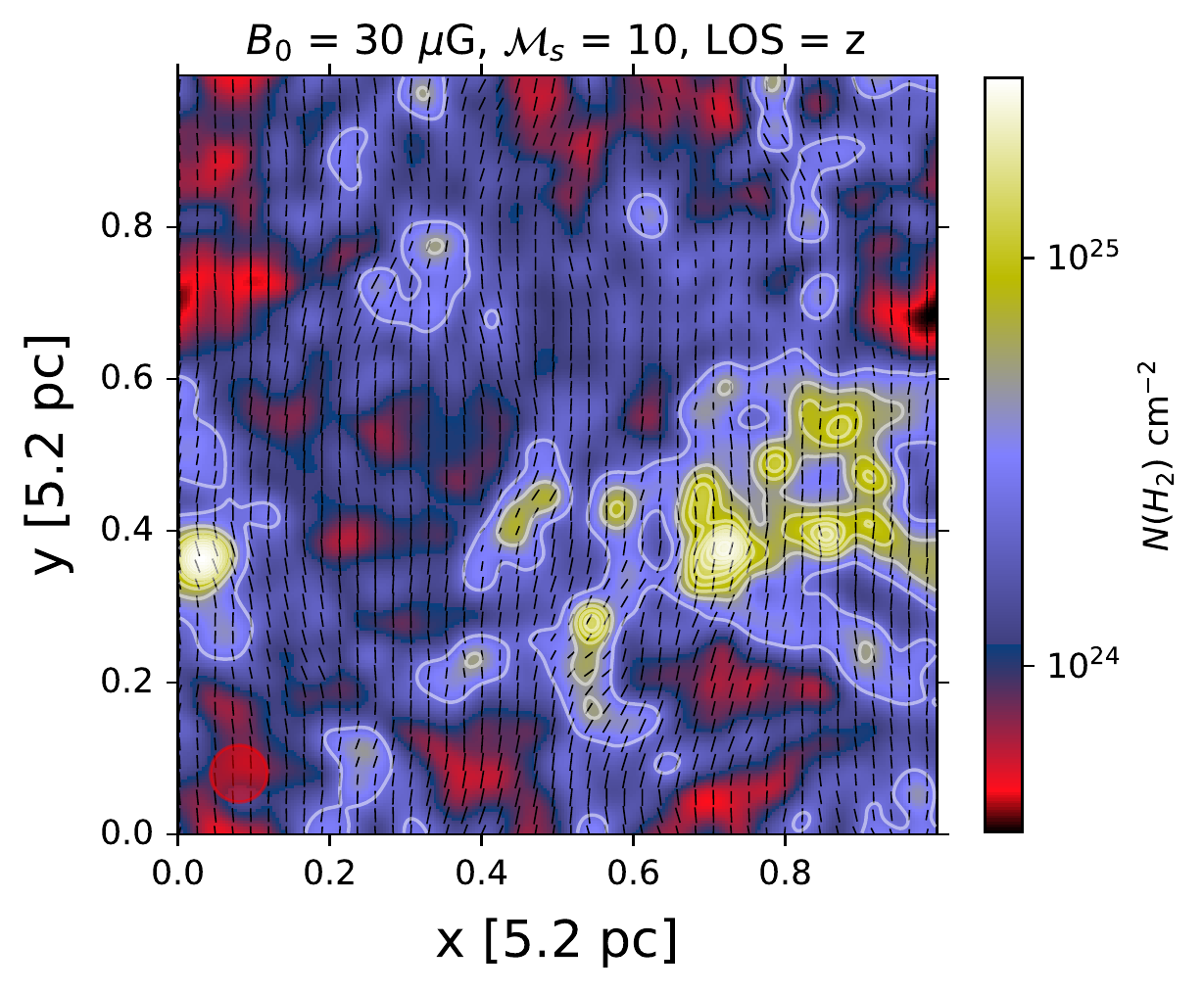}
    \includegraphics[width=0.45\textwidth]{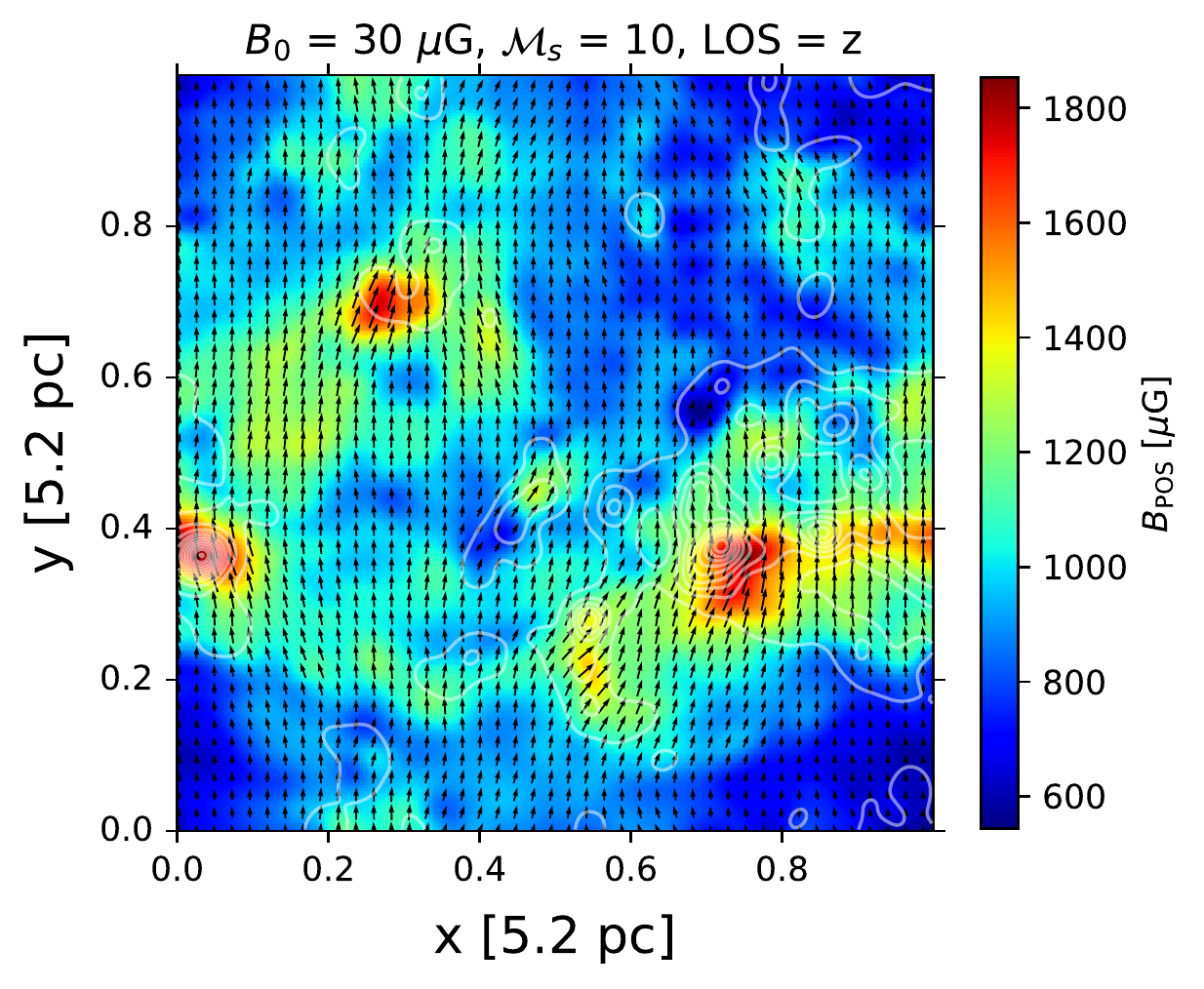}\\
    \includegraphics[width=0.45\textwidth]{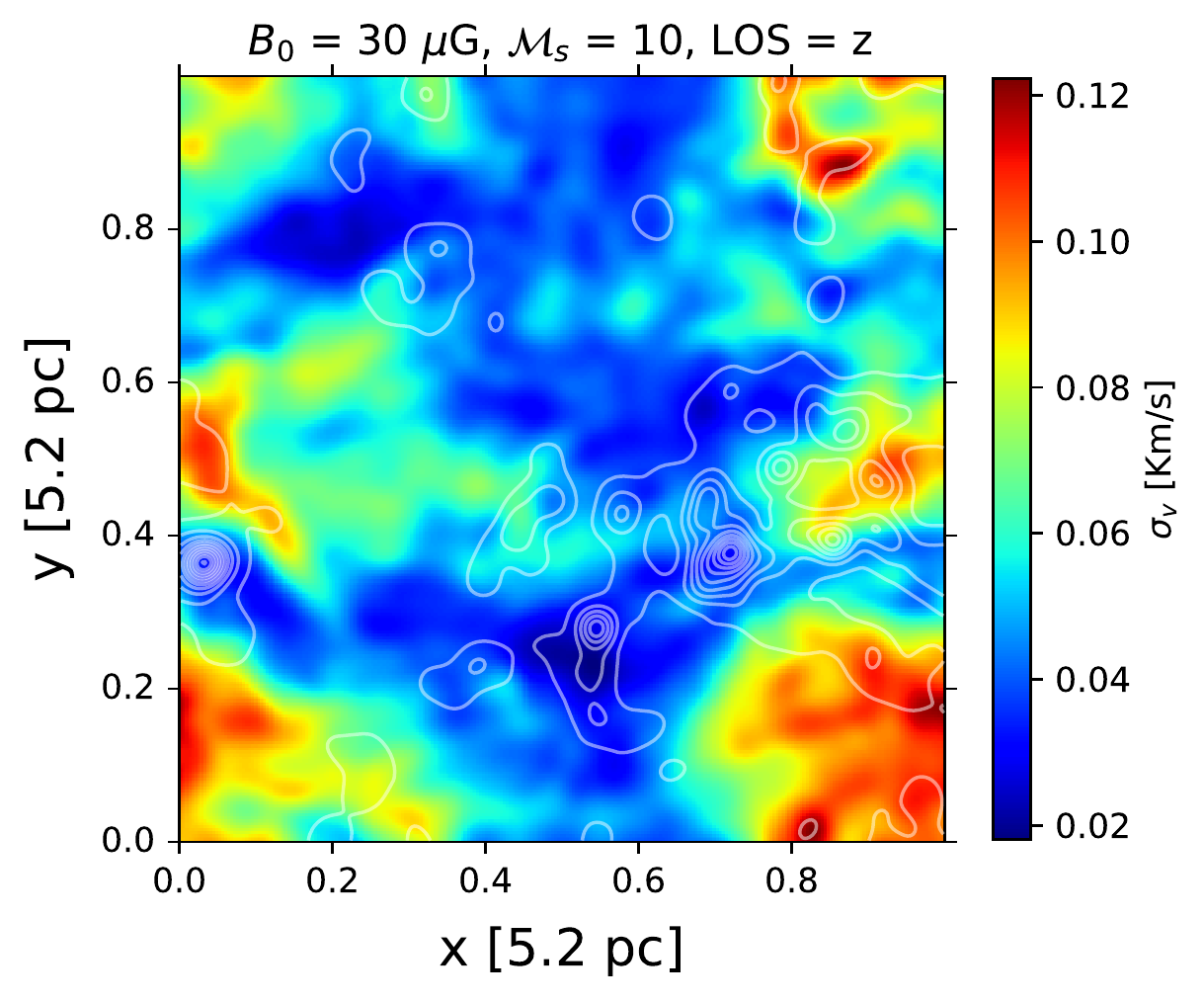}
    \includegraphics[width=0.45\textwidth]{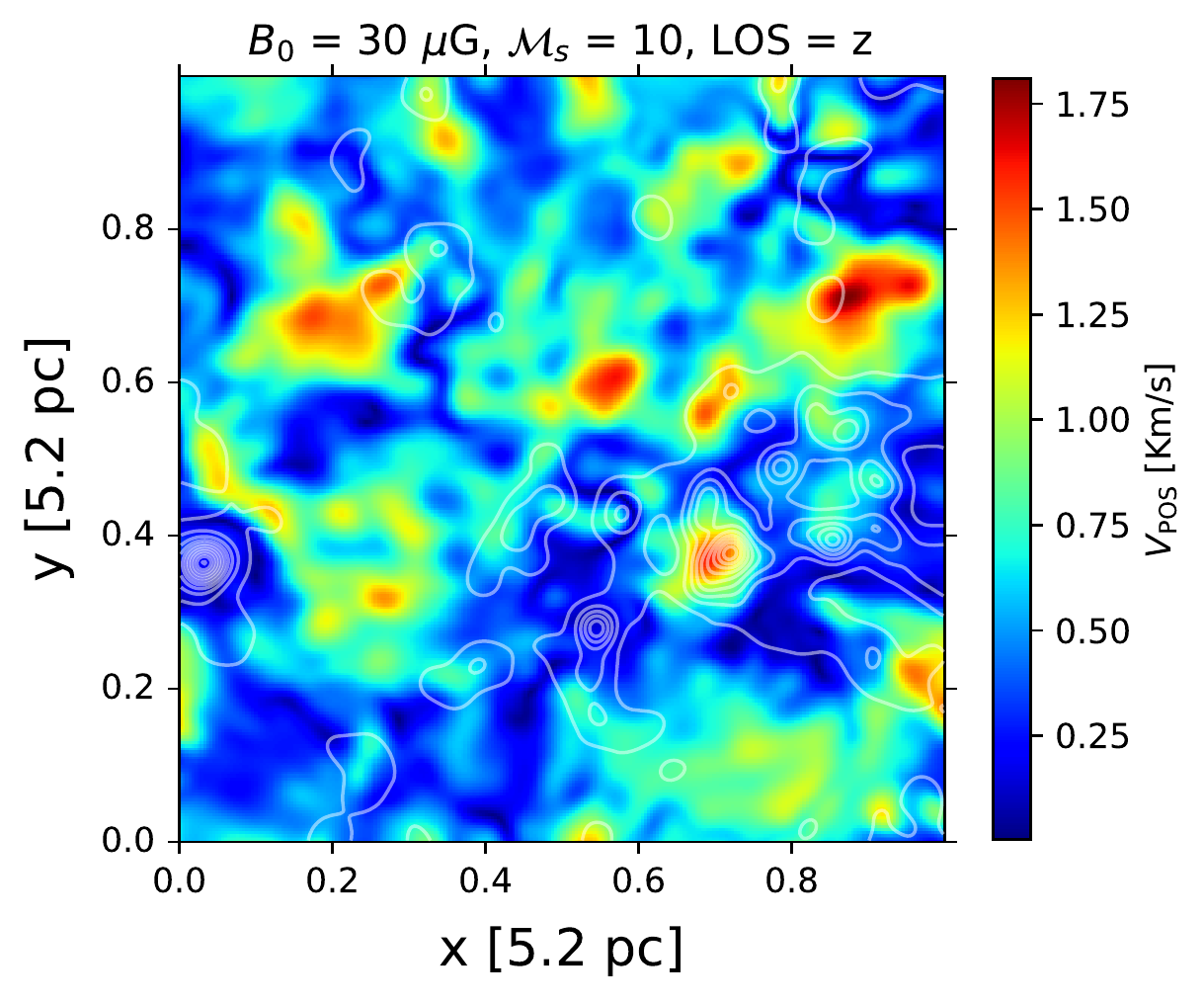}
    \caption{Projected and smoothed synthetic observations obtained from ideal MHD numerical simulations performed with the AREPO code obtained from CATS. Two-dimensional maps correspond to POS coordinates and are calculated from snapshot cubes by integration along the LOS. Parameters describing the simulations are shown in Table \ref{tab:table1}. {\it Top, Left:} synthetic B-field orientation on top of the column density; the line length is proportional to the polarization fraction, with a maximum value of 15\%. {\it Top, Right:} POS B-field; {\it Bottom, Left:} velocity dispersion; {\it Bottom, Right:} POS velocity. White contours in all panels correspond to levels of column density.}
    \label{fig:Synth_data}
\end{figure}

Before applying the shear-flow modified DCF approach (Eq.~\ref{eq:mod_dcf}) to observations, we test its accuracy using synthetic polarization data (Figure \ref{fig:Synth_data}). We use MHD simulations from the Catalogue of Astrophysical Turbulence Simulations \citep[CATS\footnote{CATS can be found at \url{https://www.mhdturbulence.com/}};][]{Burkhart2020}. CATS contains snapshots from a large number of MHD simulations performed with different codes (e.g., AREPO, ENZO, FLASH, etc.) spanning a large range of physical conditions ({\it e.g.,} low- and high-$\beta$ plasma, sub- and super-sonic gas, sub- and super-Alfv\'enic turbulence). As stated in Section \ref{sec:intro}, the motivation for our work is to estimate the B-field strength in the CND under shearing effects. Therefore, it is appropriate to select a data set with similar physical conditions to those in the CND. Table \ref{tab:table1} compiles the values of the physical variables for the CND and numerical simulations. 

\begin{table}[h!]
    \centering
    \begin{tabular}{l|c|c|c}
         & Units & CND$^{\rm a}$ & Simulation \\
         \hline
         \hline
        Magnetic field strength, $|\textbf{B}|$ & $\mu$G & 5000$^{\rm b}$ & 30 \\
        Number density, $n$ & cm$^{-3}$ & 10$^{4}$ & 10$^{3}$ \\
        Velocity Dispersion, $\sigma_{v}$ & \kms & 45 & 20 \\
        Temperature, $T$ & K & 300 & 6$^{\rm c}$ \\
        Alfv\'en Speed, $v_{A}$ & \kms & 71 & 1.7 \\
        Sound Speed, $c_{s}$ & \kms & 3.2 & 0.2 \\
        Sonic Mach Number, $\mathcal{M}_{s}$ & & 14 & 10 \\
        Alfv\'en Mach Number, $\mathcal{M}_{A}$ & & 1.1 & 1.2\\
        Plasma $\beta$ & & 0.0125 & 0.028 \\
        \hline
    \end{tabular}
    \caption{Typical values for physical conditions in the CND and corresponding values for the numerical MHD simulation used to create the synthetic dust polarization data.}
    \flushleft
    \noindent $^{\rm a}$Expected values based on literature shown in Section \ref{sec:intro}. These values will be revisited later in this work (Table \ref{tab:CND_res}).\\
    \noindent $^{\rm b}$ Assuming $|\textbf{B}|= B_{\rm POS}$.\\
    \noindent $^{\rm c}$ Derived from the value of $c_{s}$ assuming an ideal gas state equation.
    \label{tab:table1}
\end{table}

Although values of the physical variables are different, the dimensionless parameters ($\mathcal{M}_{s}$, $\mathcal{M}_{A}$, $\beta$) are very similar. The selected simulation was computed with the AREPO code \citep{2020AREPO}, and it was designed to study star-forming cores in conditions similar to those in our Galaxy \citep{Mocz2017}. The data set corresponds to a snapshot of the physical variables in a data cube with a regular grid of 256$^{3}$ voxels. The data file is in a \texttt{HDF5} format and was downloaded from the CATS website-linked repository\footnote{Data cube used for the synthetic observations \url{https://users.flatironinstitute.org/~bburkhart/data/CATS/arepo/mhd256GB30M10/snap\_073.hdf5}}. \texttt{HDF5} files are handled with the \textsc{python} package \texttt{yt} \citep{Turk2011T}. 

Using the data cube, we estimate the maps of Stokes parameters $I, Q$, and $U$ by integrating along the LOS at several rotations of the cube. The  Stokes $I$, $Q$, $U$ are computed as

\begin{eqnarray}
    I & = & \int n \left[ 1 - p_{\rm 0}\left( \frac{B^{2}_{1}+B^{2}_{2}}{B^{2}} - \frac{2}{3}\right)\right]ds, \\
    Q & = & \int p_{\rm 0} n \left( \frac{B^{2}_{2}-B^{2}_{1}}{B^{2}}\right)ds,\\
    U & = & \int p_{\rm 0} n \left( 2\frac{B_{1}B_{2}}{B^{2}}\right)ds.
    \label{eq:Stokes}
\end{eqnarray}

\noindent
\citep[e.g.,][]{WK1990,Fiege2000,PlanckIntermediateXX_2015,Chen2016}, where $B_{1},B_{2},$ and $B_{3}$ correspond to the components of the B-field, $B=\sqrt{B_{1}^{2}+B_{2}^{2}+B_{3}^{2}}$ is the total B-field strength, $n$ is the number density, $p_{0}$ is the maximum polarization fraction, and $s$ is the LOS direction along which the integration is done. For example if $s=z$, then $B_{1}=B_{x}$ and $B_{1}=B_{y}$. With the maps of the Stokes $I, Q,$ and $U$ parameters, then the polarization angle ($\phi$) and fraction ($p$) can be calculated as

\begin{eqnarray}
    \phi & = & \frac{1}{2}\arctan2(U,Q), \nonumber \\
    p & = & \frac{\sqrt{Q^{2}+U^{2}}}{I}.
    \label{eq:phi_p}
\end{eqnarray}

The uncertainty associated to the Stokes parameters are modelled as

\begin{equation}
    \sigma_{X} = 0.02\bar{X}[1 + p_{\rm G}(\mu,\sigma)].
    \label{eq:uncert}
\end{equation}

\noindent
where $X \in \{I,Q,U\}$, $\bar{X}$ is the median value, and $p_{\rm G}(\mu,\sigma)$ is a two-dimensional map of the Gaussian noise with zero mean and $\sigma=1$. Values of $\sigma_{\rm I},\sigma_{\rm Q},\sigma_{\rm U}$ are propagated through Eqs. \ref{eq:phi_p} resulting in

\begin{eqnarray}
    \sigma_{\phi} & = & \frac{1}{2}\frac{\sqrt{(Q\sigma_{U})^{2} + (U\sigma_{Q})^{2}}}{U^{2} + Q^{2}}, \nonumber \\
    \sigma_{p} & = & \frac{1}{I}\left\{\frac{(Q\sigma_{U})^{2} + (U\sigma_{Q})^{2}}{U^{2} + Q^{2}} + \left[ \left( \frac{Q}{I}\right)^{2} + \left( \frac{U}{I}\right)^{2} \right]\sigma^{2}_{I}\right\}^{1/2}.
    \label{eq:unc_phi_p}
\end{eqnarray}

To evaluate and validate any of the DCF expressions presented above it is also necessary to create maps of $\sigma_{v}$ (velocity dispersion, moment 2), $N(H_{2})$ (column density), and $U_{0}$ (velocity field, moment 0), in addition to the map of $B_{\rm POS}^{\rm Model}$ (POS component of the B-field). All these maps are calculated from the data cube by density-weighted integration along the chosen LOS: $U_{0}$ is calculated from the components of the velocity in the POS, while $\sigma_{v}$ is the root-mean-square (RMS) value of the velocity component in the LOS. The Laplacian of the POS velocity field is calculated as finite difference over the scale of one pixel in each direction. The cube is projected assuming a distance of $8.3$ kpc, which results in a pixel size of $0.5\arcsec$. Finally, in order to take into account the instrumental configurations, all maps are smoothed using a 2D Gaussian profile with an angular resolution of $4.85\arcsec$ (the FWHM equivalent to HAWC+ 53-$\micron$ observations, see Section \ref{sec:pol_obs}). Figure \ref{fig:Synth_data} displays the projected and smoothed POS B-field orientation (from rotated polarization vectors) over the column density ({\it Top, Left}) and the B-field strength ({\it Top, Right}). The velocity dispersion ({\it Lower, Left}) and the POS velocity ({\it Lower, Right}) are also shown. In all maps, contours of the column density are also included for reference. 

\begin{figure}[!h]
    \centering
    \includegraphics[width=0.90\textwidth]{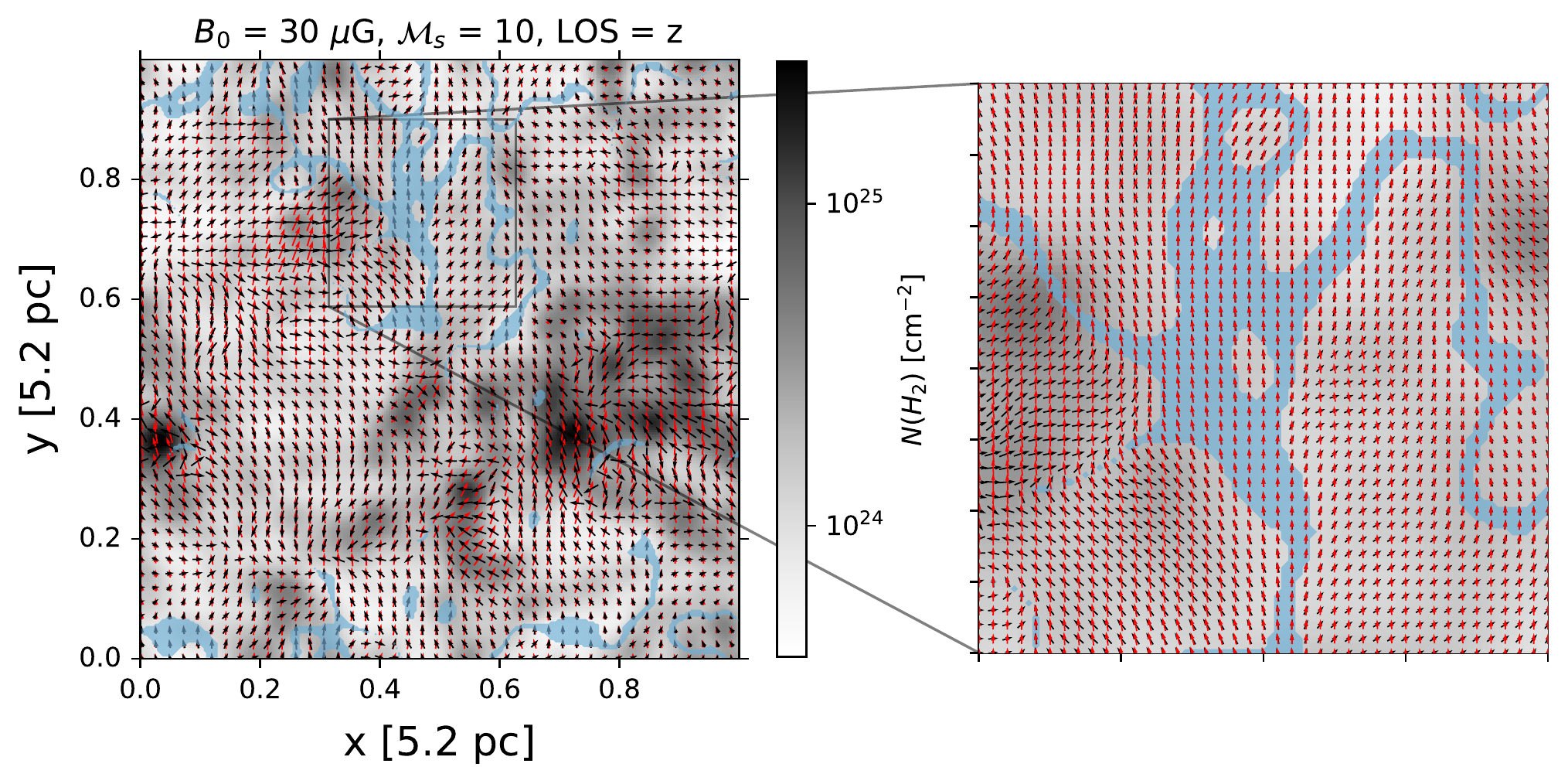}
    \caption{{\it Left:} Shear-flow mask for synthetic data analysis. Two criteria are used to create the mask (blue shaded region): 1) pixels where the magnetic field (red arrows) and flow direction (black arrows) differ by less than $\pm$10$^{\rm o}$, and 2) pixels where the gradient of the flow is different from zero. {\it Right:} Zoom into a region of the parallel flow and magnetic-field mask.}
    \label{fig:Synth_data_mask}
\end{figure}

As Eq.~\ref{eq:mod_dcf} is derived for the case where the B-field is primarily aligned with the steady-state, sheared flow, we want to apply this expression only to locations of the polarimetric map where this condition is met. We ensure this criteria by selecting only pixels where: 1) the difference between the model POS magnetic field vectors and velocity vectors differs by less than $\pm$10$^{\rm o}$, and 2) the gradient of the POS velocity field is nonzero. In Section \ref{sec:pol_obs}, we will select polarization measurements with a signal-noise-radio $\ge3$ in the polarization fraction, which have an associated angular uncertainty of $\le9.6^{\circ}$. Thus, the first condition ensures that all selected polarization measurements in the simulation have similar angular dispersion to the observations. The regions where these criteria are met are shown in Figure \ref{fig:Synth_data_mask} as blue shaded areas in a map of both the POS B-field (red arrows) and velocity field (black arrows). From this point forward, such pixels constitute the shear-flow mask, which will be used for all synthetic variables involved in the DCF calculations.

We estimate the angular dispersion inside the shear-flow mask in the synthetic polarimetric data map using the methodology presented by \citet{Hildebrand2009,Houde2009}. We describe the details of this technique in Appendix \ref{sec:app_a} and show the computed parameters in Table \ref{tab:res_synth} (rows 2, 5 - 7). For the rest of the physical variables (e.g., column density, velocity dispersion, large-scale, shear flow, and POS magnetic field strength), median values within the mask are calculated and used for the subsequent DCF calculations. Since the MHD simulations do not provide estimation of uncertainties for physical variables, only the uncertainties of parameters from the dispersion analysis are reported here.

\begin{table}
\begin{tabular}{c|c|c|c|c}
    & LOS & $x$ & $y$ & $z$ \\
    \hline
    \hline
    1 & $N(\rm H_{2})$ [cm$^{-2}$] & 1.95$\times$10$^{24}$ & 1.89$\pm\times$10$^{24}$ & 1.27$\times$10$^{24}$\\
    2 & $\Delta^{\prime}$ [cm] & (0.76$\pm$0.09)$\times$10$^{17}$ & (1.78$\pm$0.21)$\times$10$^{17}$ & (0.73$\pm$0.09)$\times$10$^{17}$ \\
    3 & $\rho$ [g cm$^{-3}$] & (1.54$\pm$0.05)$\times$10$^{-16}$ & (6.51$\pm$0.01)$\times$10$^{-17}$ & (9.43$\pm$0.02)$\times$10$^{-17}$ \\
    4 & $\sigma_{\rm v}$ [km s$^{-1}$] & 0.07 & 0.07 & 0.06 \\
    \hline
    5 & $\sigma_{\phi}$ & 0.31$\pm$0.03 & 0.10$\pm$0.03 & 0.10$\pm$0.03 \\
    6 & $\delta$ [arcsec] & 5.42$\pm$0.19 & 8.61$\pm$0.07 & 11.37$\pm$0.10\\
    7 & $\mathcal{N}$ & 0.12$\pm$0.01 & 0.15$\pm$0.01 & 0.04$\pm$0.01\\
    \hline
    8 & $U_{\rm 0}$ [km s$^{-1}$] & 0.02 -- 2.69 & 0.00 -- 1.64 & 0.01 -- 1.22 \\
    9 & $\nabla^{2}(U_{\rm 0})$ [km s$^{-1}$] & -0.05 -- 0.20 & -0.04 -- 0.17 & -0.05 -- 0.17 \\
    \hline
    10 & $B_{\rm POS}^{\rm Model}$ [$\mu$G] & 123 & 1005 & 925 \\
    \hline
    11 & $B_{\rm POS}^{\rm DCF}$ [$\mu$G] & 862$\pm$51($>$100\%) & 1931$\pm$116(81 -- 39\%) & 1613$\pm$131(60 -- 89\%) \\
    12 & $B_{\rm POS}^{\rm DCF,F}$ [$\mu$G] & 3927$\pm$244($>$100\%) & 534$\pm$35(38 -- 45\%) & 577$\pm$89(28 -- 47\%)\\
    13 & $B_{\rm POS}^{\rm DCF,SF}$ [$\mu$G] & 871$\pm$52($>$100\%) & 1003$\pm$61(6\%) & 840$\pm$114(3 -- 22\%)\\
    \hline
\end{tabular}
\caption{POS B-field strengths calculated using synthetic polarimetric data for various sight lines. Rows 1-4 contain the values of column density, clouds' effective depth, mass density, and velocity dispersion. Rows 5-7 correspond to the results of the dispersion analysis. Rows 8 and 9 are the values of the gas flow. Row 10 is the true value of the POS B-field strength, $B_{\rm POS}^{\rm Model}$. Rows 11 - 13 display the three values of the estimated B-field strength: 1) Classical DCF ($B_{\rm POS}^{\rm DCF}$, Eq. \ref{eq:DCF_og}), 2) Large-scale flow DCF approximation ($B_{\rm POS}^{\rm DCF,F}$, Eq. \ref{eq:DCF_lsflow}), and 3) Shear-flow DCF approximation ($B_{\rm POS}^{\rm DCF,SF}$, Eq. \ref{eq:mod_dcf}).}
\label{tab:res_synth}
\end{table}

Table \ref{tab:res_synth} shows the results obtained for all three different LOS: $x, y, z$ (2nd, 3rd, and 4th columns, respectively). Rows 1 - 4 display the values of the physical variables needed for all three estimates of the B-field strengths: mass density (gas column density and depth of the cloud) and velocity dispersion. Rows 5 - 7 are the results of the dispersion analysis (Appendix \ref{sec:app_a}). Although only $\sigma_{\phi}$ is needed for the DCF calculation, $\delta$ and $\mathcal{N}$ are important to determine if the effect of the turbulence in the gas/B-field is fully resolved by the polarimetric data. As shown, $\delta>\sqrt{2}FWHM$ of the synthetic beam, thus the turbulence length scale is resolved in the synthetic polarimetric observations along the shear flow. Rows 8 and 9 correspond to the minimum and maximum values of the large-scale flow and its Laplacian, which are necessary for the modified DCF estimates.  The true POS magnetic field from the model, $B_{\rm POS}^{\rm Model}$, is in Row 10. We calculate three values of the POS B-field strength: 1) Classical DCF ($B_{\rm POS}^{\rm DCF}$, Eq. \ref{eq:DCF_og}), 2) Large-scale flow DCF approximation ($B_{\rm POS}^{\rm DCF,F}$, Eq. \ref{eq:DCF_lsflow}), and 3) Shear-flow DCF approximation ($B_{\rm POS}^{\rm DCF,SF}$, Eq. \ref{eq:mod_dcf}). These values and their uncertainties are shown in rows 11 - 13. DCF estimates correspond to median values calculated over the shear-flow mask and their uncertainties reflects how the individual uncertainties in all variables affect the median (see Appendix \ref{sec:unc}). Percentage difference between the model value, $B_{\rm POS}^{\rm Model}$, and the DCF estimates are shown in parenthesis in rows 11-13.

The $x$ LOS displays the largest percentage difference for all three B-field strengths (in comparison to the $y$ and $z$ LOS). This result is due to the fact that the MHD simulations had their initial B-field orientation along the $x-$direction \citep{Mocz2017}. Therefore, any DCF type approximation will fail if the B-field is too close to the LOS \citep{Houde2004}. 

For the other two LOS ($y, z$), the classical DCF approximation provides values $\sim$40-90\% larger than $B_{\rm POS}^{\rm Model}$. The classical DCF method typically overestimates the B-field strength in the POS by a factor 0.2-0.7 \citep[i.e. discussion in Section 2.2.3 by ][]{Skalidis2021}. However, the angular dispersion analysis (Appendix \ref{sec:app_a}) should have corrected for this overestimation. This result indicates that the classical DCF method with the angular dispersion correction still overestimates the B-field strength by at least a factor of $\sim1.4$ for the shear flow regions in the synthetic observations. The large-scale flow modification $B_{\rm POS}^{\rm DCF,F}$, on the other hand, produces values in all LOS analyses with percentage errors comparable to those the classical DCF but systematically producing values under $B_{\rm POS}^{\rm Model}$. This particular difference might result from the the fact that our chosen mask includes regions of large velocity differences, thus implying the importance of considering the Laplacian of $U_{0}$ as in Eq.~\ref{eq:mod_dcf}. Finally, the shear-flow correction, $B_{\rm POS}^{\rm DCF,SF}$, provides the B-field strength closest to the true values. We estimate that the shear-flow correction, $B_{\rm POS}^{\rm DCF,SF}$, measures the true value of the B-field with a relative deviation as low as 3\% and as large as 22\%. 

Although these results suggest that incorporating terms related to structured flows ($U_{0},\nabla^{2}U_{0}$) into the DCF approximation improves the accuracy in calculating the POS B-field strength, it is important to keep in mind that this trend might not extend to simulations with different initial physical condition from those assumed here. Here, we show that the appropriate DCF method must be based on the physical conditions of the astrophysical environment.

\section{Application to the Galactic's Circum-Nuclear Disk}\label{sec:CND}

The B-field in the CND has been studied in FIR, millimeter, and radio observations. Polarized dust observations (100, 850 $\micron$) indicate that the B-field orientation in the CND displays an axisymmetric configuration consistent with a self-similar disk model \citep[i.e.,][]{Hildebrand1993,Hsieh2018}. These results have assumed an \textit{a\ priory} spiral and/or toroidal B-field configuration. Here, we characterize the POS B-field orientation and strength from a model-independent approach. We use the gas kinematics to separate the gas streamers toward Sgr~A$^{*}$ and several DCF approximations to quantify the B-field strength and energy balance of these streamers.

\subsection{Far-Infrared Polarimetric Observations}\label{sec:pol_obs}

\begin{figure}[h!]
    \centering
    \includegraphics[width=\textwidth]{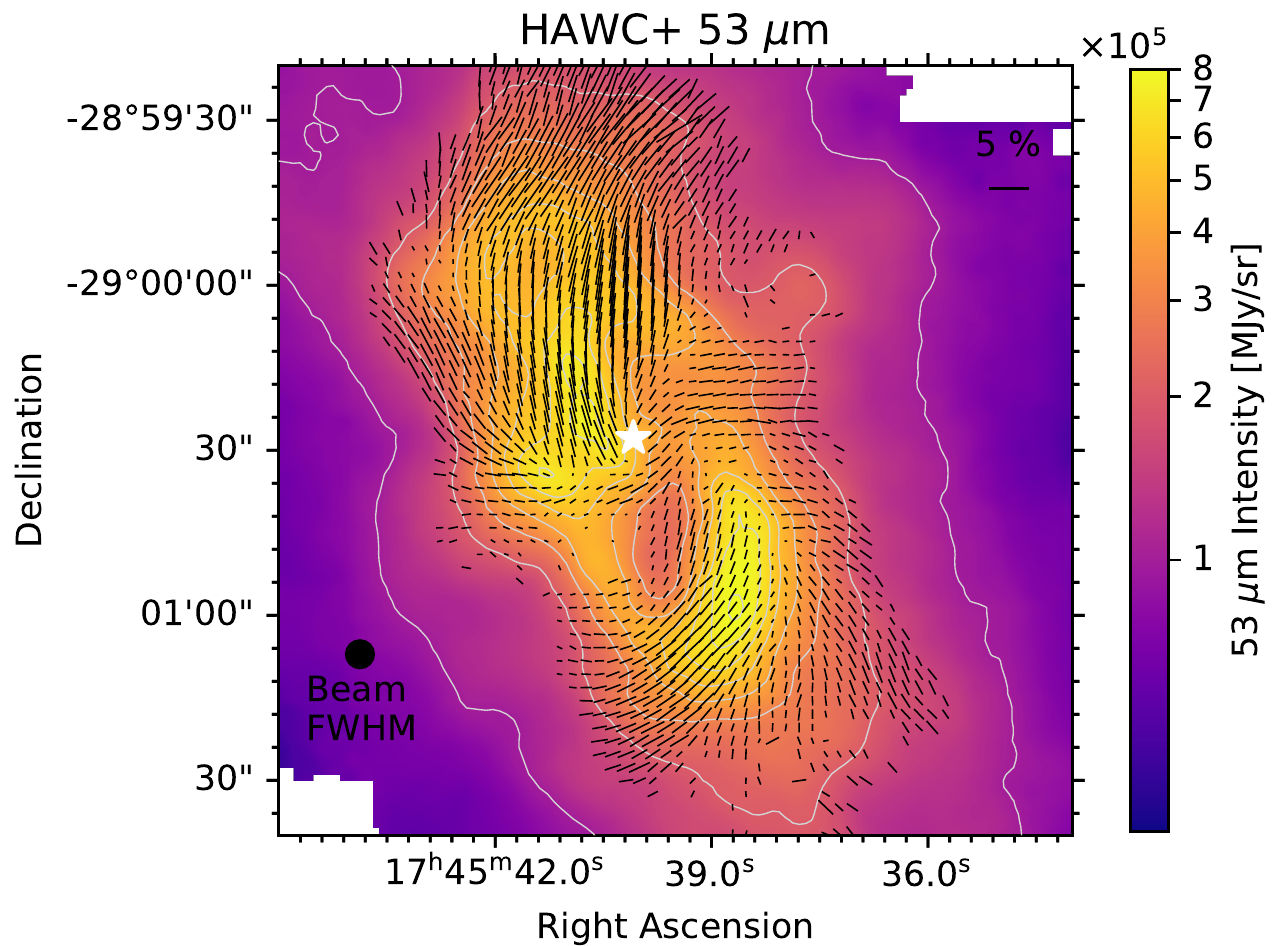}
    \caption{SOFIA/HAWC+ 53-$\micron$ observations of the CND. B-field pseudovectors (black) representing the orientation of the magnetic field are plotted on the 53-$\micron$ intensity map (Stokes $I$). The location of Sgr A* is depicted by the white star and the beam size is displayed in the lower left corner. Vectors displayed satisfy the condition $p/\sigma_{p}\ge3$. A legend with a 5\% polarization fraction is shown.}
    \label{fig:pol_data}
\end{figure}

We use publicly available SOFIA/HAWC+ polarimetric observations of the CND obtained under proposal ID 70\_0509 (Guaranteed Time Observations by the HAWC+ Team, PI: Dowell, C.D.). Continuum intensity maps of the Stokes parameter $I, Q, U$ were obtained at 53 $\micron$ using the standard Chop Nod Match Chop (CNM) observing mode. These maps have an angular resolution of $4.85$\arcsec, which correspond to the FWHM value of the beam. From the Stokes parameter maps, the polarization angles and polarization fraction maps are calculated as $\phi=0.5\arctan2(U,Q)$ and $p_{\rm m}=\sqrt{Q^{2}+U^{2}}/I$. According to this definition, $-90^{\rm o}<\phi<90^{\rm o}$, measured east of north in a counterclockwise direction. In addition, polarization fractions are debiased using $p = \sqrt{p^{2}_{\rm m} - \sigma^{2}_{p}}$ \citep{Serkowski1974}, where $\sigma_{p}$ is the uncertainty of the measured polarization fraction, $p_{\rm m}$. The resulting polarimetric data for the CND are shown in Figure \ref{fig:pol_data}. The inferred POS B-field orientations are displayed on top of the 53-$\micron$ total intensity. These B-field orientations are determined by rotating the polarization angles (E-vector) $\phi$ by 90$^{\rm o}$. The length of the B-field orientation is proportional to the value of polarization fraction. For reference, the length corresponding to 5\% is shown as well as the position of Sgr~A$^{*}$. Only the polarization measurements that satisfied the conditions of $p/\sigma_{p}\ge3$, $p < 50\%$, and $I \ge 1.5\times$10$^{5}$ MJy/sr are displayed in Figure \ref{fig:pol_data}.

\subsection{Velocity Field}

Three streamers had been identified in the CND using observations of ionized gas:  {\it Northern Arm}, {\it Eastern Arm}, and {\it Western Arc}. \citet{Zhao2009} studied these ionized gas streamers in the CND using data from the Very Large Array. They characterized the kinematics of the streamers using the proper motions of compact HII regions at two epochs and reported the best-fit parameters for each streamer assuming the gas is moving in partial Keplerian orbits around Sgr~A$^{*}$. We use the POS projection of these orbital velocities in the streamers to determine the velocity field in the shear flow, $U_{0}(x,y)$, which is necessary for evaluating the modified DCF (Eq. \ref{eq:mod_dcf}) approximation.

\begin{figure}[h!]
    \centering
    \includegraphics[width=\textwidth]{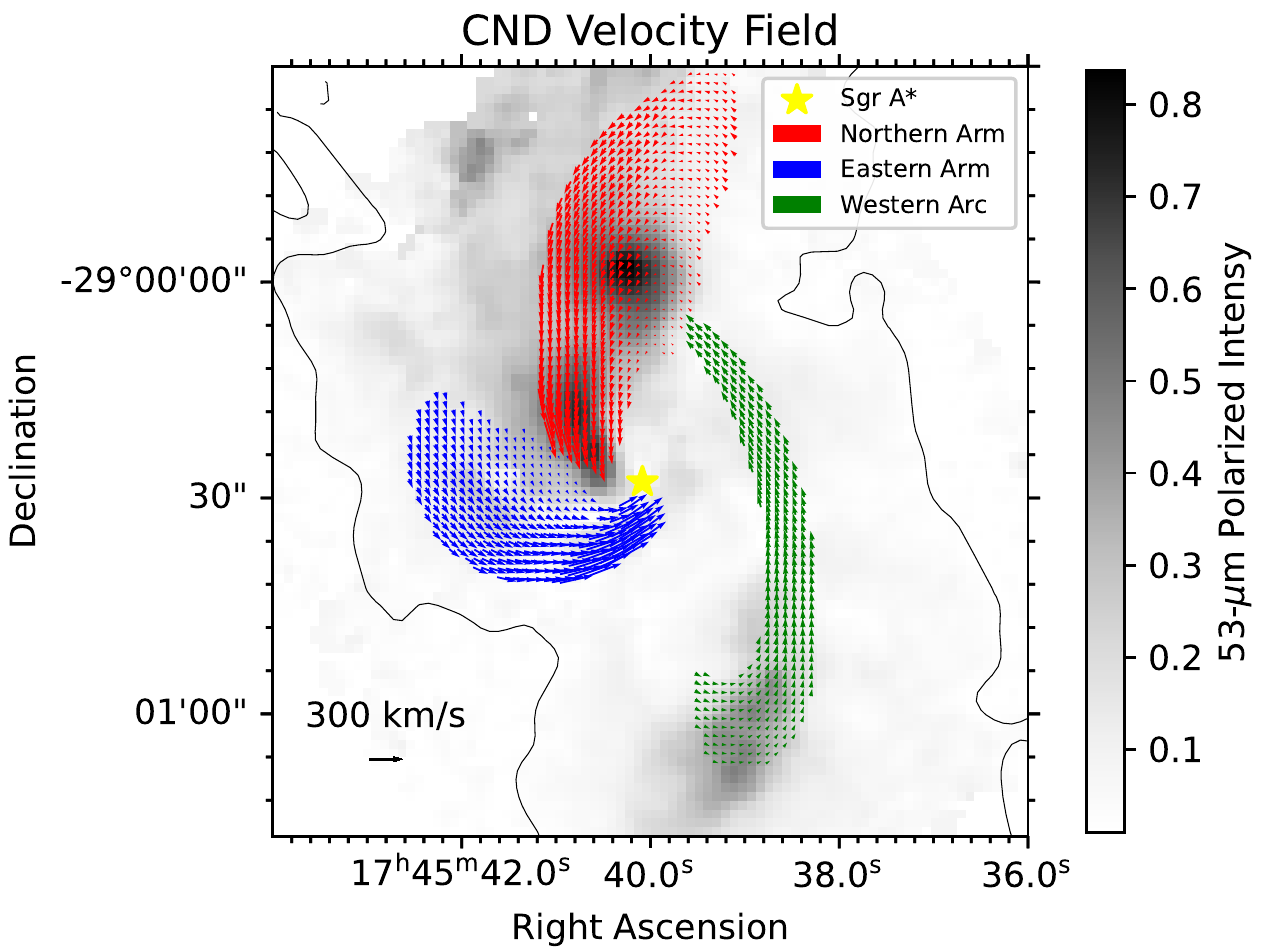}
    \caption{POS velocity field for the CND ionized streamers {\it Northern Arm} (red), {\it Eastern Arm} (blue), and {\it Western Arc} (green). Velocities are derived from Keplerian-orbit modelling with parameters in Table \ref{tab:orbits} \citep{Zhao2009}. The yellow star represents the location of Sgr A*. Background image corresponds to the HAWC+ 53-$\micron$ polarized intensity. Black lines shows the area where the condition $PI/\sigma_{p_{I}}\ge3$.}
    \label{fig:Velocity_field}
\end{figure}

We calculate the three-dimensional velocity vector for each orbit using the function \texttt{KeplerEllipse} from the \textsc{python} package \texttt{pyAstronomy}\footnote{\texttt{pyAstronomy} can be found at \url{https://pyastronomy.readthedocs.io/en/latest/}} \citep{pya}. We use the parameters reported by \citet{Zhao2009}, reproduced here in Table \ref{tab:orbits}. For each streamer, a total of 100 individual orbits are calculated. This was done by choosing values of eccentricity ($e$) and semimajor axis ($a$) uniformly distributed in the range [$e-\Delta e$, $e+\Delta e$] and [$a-\Delta a$, $a+\Delta a$] while keeping all other parameters fixed. However, because the large fractional error in the value of the argument of the perifocus ($\omega$) in the western arc (218\%), this parameter was also varied for this case. Changing the values of the other parameters showed no effect on the kinematic variables. Each orbit is initially evaluated 350 temporal points during a complete orbital period; this, of course, produces an entire ellipse. We then limit the bundle to the extend of the streamer (as seen in Figure \ref{fig:Velocity_field}) by applying cuts based on a) a SNR in the polarized intensity of $pI/\sigma_{pI}\ge3$, and b) the projected azimuthal angle (measured east of north) with respect to the position of Sgr A*. These azimuthal cuts were applied at -10$^{\circ}$ and 60$^{\circ}$ for the northern arm, 65$^{\circ}$ and -135$^{\circ}$ for the eastern arm, and -10$^{\circ}$ and -170$^{\circ}$ for the western arc. 

The results from evaluating the \texttt{KeplerEllipse} function are the velocity components $v_{x}, v_{y}, v_{z}$ and their corresponding coordinates $x,y,z$ (with respect to the position of Sgr~A$^{*}$). Using this information, the POS velocity components ($v_{x},v_{y}$) were mapped onto the same grid as the HAWC+ polarimetric data. All velocity values within the HAWC+ pixel scale are averaged, and the POS velocity is calculated as $v_{\rm POS}=\sqrt{v_{x}^{2} + v_{y}^{2}}$. Figure \ref{fig:Velocity_field} displays the POS velocity vector field for all three streamers overlaid to the HAWC+ 53-\micron\ polarized intensity, $PI$. POS velocities ranges are $\sim$20 - 326 \kms~for the northern arm, $\sim$30 - 432 \kms~for the eastern arm, and $\sim$44 - 119 \kms~for the western arc. In creating these velocity fields, we retained only pixels that show $PI/\sigma_{p_{I}}\ge3$.

\begin{table}
\centering
\begin{tabular}{lcccc}
    \hline
    \hline
    Orbital Parameters & Units & Northern Arm & Eastern Arm & Western Arc \\
    \hline
    Eccentricity ($e$) &    & 0.83$\pm$0.10 & 0.82$\pm$0.05 & 0.20$\pm$0.15 \\
    Semimajor axis ($a$) & pc & 0.99$\pm$0.44 & 1.40$\pm$0.68 & 1.11$\pm$0.06 \\
    Longitude of the ascending node ($\Omega$) & deg & 64$\pm$28 & -42$\pm$11 & 71$\pm$6 \\
    Argument of perifocus ($\omega$) & deg & 132$\pm$40 & -280$\pm$8 & 22$\pm$48 \\
    Inclination ($i$) & deg & 139$\pm$10 & 122$\pm$5 & 117$\pm$3 \\
    Perifocal distance ($q$) & pc & 0.17 & 0.25 & 0.89 \\
    Period (T) & 10$^{3}$ yr & 45 & 76 & 54 \\
    \hline
\end{tabular}
\caption{Orbital parameters of the gas streamers around Sgr~A$^{*}$. Values were reported by \citet{Zhao2009} and orbits were created using the python package \texttt{pyAstronomy}.}
\label{tab:orbits}
\end{table}

\subsection{Column Density and Velocity Dispersion}\label{sec:nh2_sigmav}

\begin{figure}[h!]
    \centering
    \includegraphics[width=0.49\textwidth]{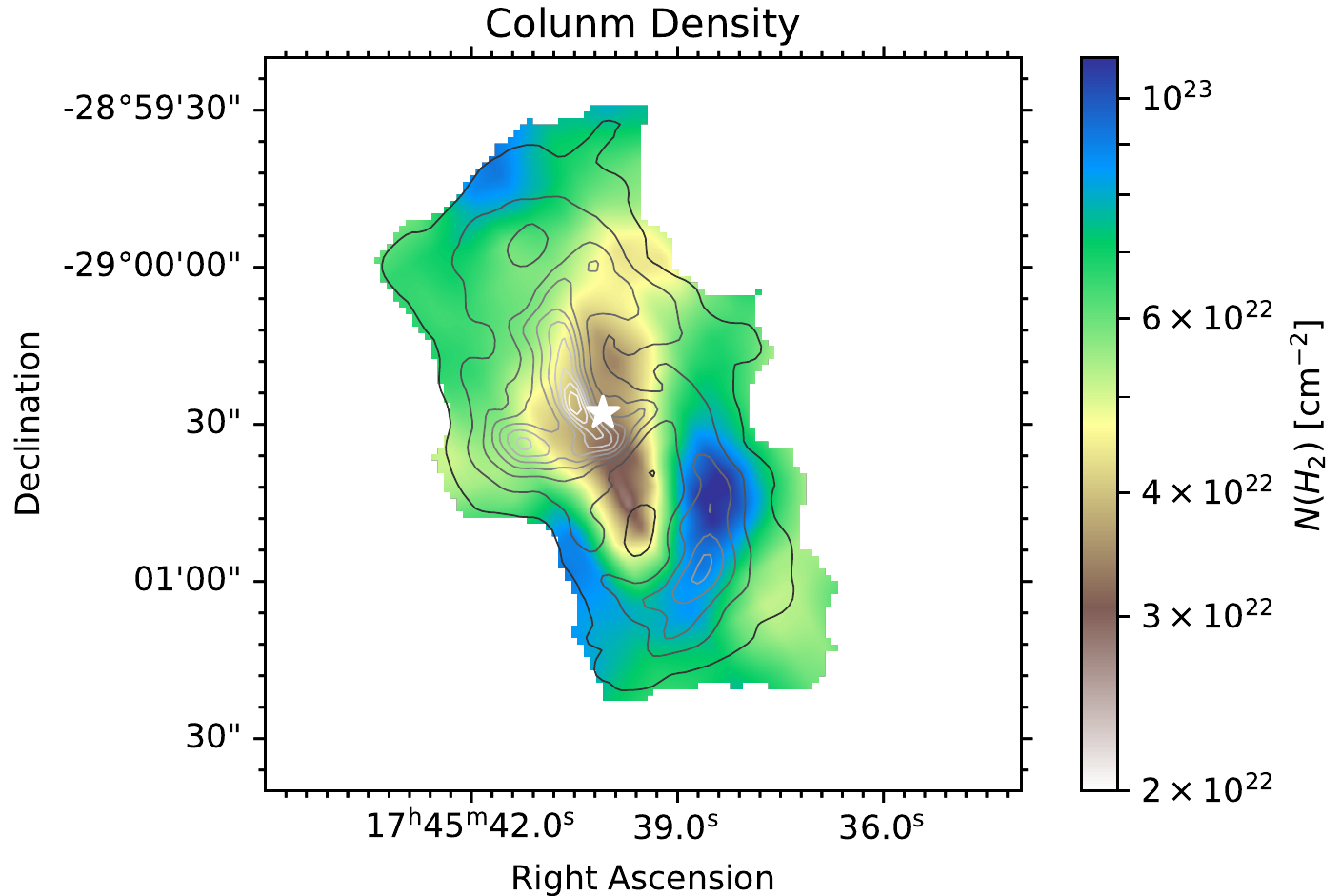}
    \includegraphics[width=0.45\textwidth]{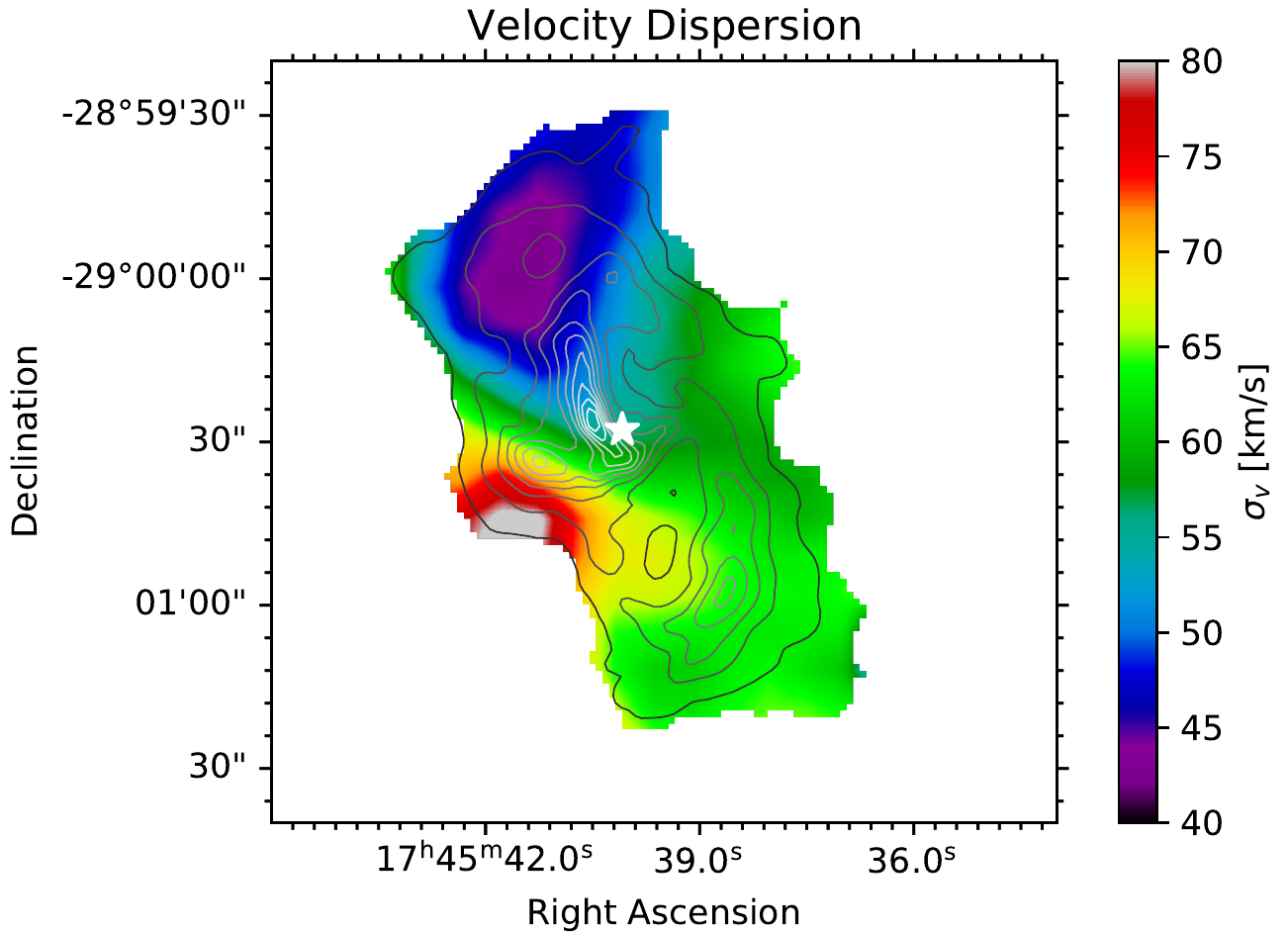}
    \caption{Auxiliary data for DCF calculations. {\it Left:} Column density map for the CND, calculated via SED fitting of \textit{Herschel} data. {\it Right:} Velocity dispersion map calculated from [CII] spectral emission observed with SOFIA/GREAT. In both panels, contours correspond to the HAWC+ 53-$\micron$ intensity levels and the white star represents the location of Sgr A*.}
    \label{fig:aux_data}
\end{figure}

Two other physical variables necessary to estimate the POS B-field strength are the column density, $N(H_{2})$, and LOS velocity dispersion, $\sigma_{v}$. Column density maps are typically constructed by means of Spectral Energy Distribution (SED) fittings using multiple wavelength measurements in the range of FIR to mm \cite[e.g.][]{Chuss2019}. We use column density values for the CND from the publicly available maps\footnote{Map can be found in \url{http://www.astro.cardiff.ac.uk/research/ViaLactea/}} created by the HiGAL project \citep{Molinari2010}. These maps were calculated using multi-temperature differential SED fitting of \textit{Herschel} wavelengths: 70, 140, 250, 350, 500 $\micron$ \citep{Marsh2015,Marsh2017}. Figure \ref{fig:aux_data}-left displays the column density map for the CND. The values of $N(H_{2})$ range from $\sim10^{22}$ cm$^{-2}$ closer to the location of Sgr A* to $\sim10^{23}$ cm$^{-2}$ in the western arc. This map has an angular resolution of $14$\arcsec.

The velocity dispersion map (Figure \ref{fig:aux_data}, {\it Right}) was obtained from the line widths of [CII] spectral observations taken with SOFIA/GREAT \citep{2012HeyminckGREAT}. The [CII] spectral cube has spatial dimensions of 30$\times$45 pixels and covers velocity channels between -220 and 420 km s$^{-1}$ with a spectral resolution of 385 m s$^{-1}$. The angular resolution of this data cube is $14.1$\arcsec~and a pixel scale of 7$^{\prime\prime}$. Figure \ref{fig:aux_data}({\it Right}) shows the resulting velocity dispersion map. Dispersion values are calculated from the line widths and then corrected to disregard the linewidth contribution due to thermal motions of the gas/dust. Values range between $\sim$40 \kms~close to the edge of the northern arm, to $\sim$80 \kms~in the Eastern Arm. Full details on the calculation of this map can be found in Appendix \ref{sec:app_b}.

\subsection{Magnetic Field Strength}\label{sec:b_res}

We now have all the necessary data to estimate the B-field strength using the modified DCF. First, we calculate the strength for the entire CND region. All variable values and results are displayed in Table \ref{tab:CND_res}, column 2. We want to focus only in the ``dense" and brightest regions, with the highest SNR, as detected by the HAWC+ observations. Thus, we apply a cutoff in Stokes $I$ intensity at $4$ Jy/pixel (1.5$\times$10$^{5}$ MJy/sr) -- same as shown in Figure \ref{fig:pol_data}. The dispersion analysis described in the Appendix \ref{sec:app_a} was applied to all polarization measurements above the intensity level that satisfies the 3$\sigma$ criteria in polarization fraction. We obtained values of $\sigma_{\phi} = 0.48\pm0.07$ radians and $\delta=11.25\pm0.27\arcsec$. Since $\delta > \sqrt{2}W = 2.90\arcsec$, the turbulent scale in the gas sampled by the polarimetric data is properly resolved, and therefore, the B-field strength can be calculated.

\begin{table}
\footnotesize
\hspace{-0.5in}
\begin{tabular}{c|c|c|c|c|c}
     &  & CND & Eastern Arm & Northern Arm & Western Arc \\
    \hline
    \hline
    1 & $N(H_{2})$ [cm$^{-2}$] & 6.36$\times$10$^{22}$ & 5.48$\times$10$^{22}$ & 5.45$\times$10$^{22}$ & 8.25$\times$10$^{22}$\\
    2 & $\Delta^{\prime}$ [cm] & (3.38$\pm$0.51)$\times$10$^{18}$ & (1.18$\pm$0.19)$\times$10$^{18}$ & (1.15$\pm$0.31)$\times$10$^{18}$ & (0.71$\pm$0.11)$\times$10$^{18}$ \\
    3 & $\rho$ [g cm$^{-3}$] & (0.89$\pm$0.01)$\times$10$^{-19}$ & (2.12$\pm$0.02)$\times$10$^{-19}$ & (2.21$\pm$0.02)$\times$10$^{-19}$ & (5.41$\pm$0.05)$\times$10$^{-19}$ \\
    4 & $\sigma_{v}$ [km s$^{-1}$] & 35 & 38 & 29 & 36 \\
    \hline
    5 & $\sigma_{\phi}$ [rad] & 0.48$\pm$0.07 & 0.50$\pm$0.07 & 0.14$\pm$0.07 & 0.36$\pm$0.10 \\
    6 & $\delta$ [arcsec] & 11.25$\pm$0.27 & 5.08$\pm$0.16 & 3.25$\pm$0.27 & 10.25$\pm$0.47 \\
    7 & $\mathcal{N}$ & 0.85$\pm$0.02 & 0.82$\pm$0.04 & 2.1$\pm$0.3 & 0.3$\pm$0.0 \\
    \hline
    8 & $U_{0}$ [km s$^{-1}$] & 20$\pm$3 - 432$\pm$48 & 30$\pm$1 - 432$\pm$48 & 20$\pm$3 - 326$\pm$10 & 44$\pm$5 - 119$\pm$5 \\
    9 & $\nabla^{2}(U_{0})$ [km s$^{-1}$]$^{a}$ & (-8 -- 9)$\times10^{3}$ & (-1 -- 1)$\times10^{3}$ & (-10 -- 6)$\times10^{2}$ & -50 -- 62 \\
    \hline
    10 & $B_{\rm POS}^{\rm DCF}$ [mG] & 7.3$\pm$0.6 & 12.9$\pm$1.1 & 33.3$\pm$5.9 & 24.0$\pm$2.1 \\
     &  & (5.4$\pm$0.4 - 9.9$\pm$0.8) & (9.1$\pm$0.8 - 15.2$\pm$1.3) & (30.9$\pm$5.5 - 40.3$\pm$7.2) & (17.0$\pm$1.5 - 27.4$\pm$2.3) \\
    11 & $B_{\rm POS}^{\rm DCF,F}$ [mG] & 2.9$\pm$0.9 & 3.1$\pm$1.4 & 19.7$\pm$6.1 & 4.5$\pm$1.8 \\
     &  & (0.3$\pm$1.4 - 17.3$\pm$3.2) & (0.4$\pm$2.1 - 37.4$\pm$7.6) & (2.9$\pm$6.0 - 32.2$\pm$6.4) & (0.7$\pm$1.5 - 13.2$\pm$2.0) \\
    12 & $B_{\rm POS}^{\rm DCF,SF}$ [mG] & 1.0$\pm$0.2 & 4.0$\pm$1.2 & 5.1$\pm$0.8 & 8.5$\pm$2.3 \\
    &  & (0.3$\pm$0.5 - 2.9$\pm$0.8) & (0.9$\pm$1.7 - 13.4$\pm$3.8) & (1.3$\pm$1.0 - 15.6$\pm$2.8) & (2.7$\pm$2.6 - 26.5$\pm$3.6) \\
    \hline
\end{tabular}
\caption{POS magnetic field strength calculated for the overall CND and streamers. For each streamer, three values are displayed: 1) classical DCF approximation ($B_{\rm POS}^{\rm DCF}$), 2) large-scale flow modified DCF approximation ($B_{\rm POS}^{\rm DCF,F}$), and 3) shear-flow modified DCF approximation ($B_{\rm POS}^{\rm DCF,SF}$). Rows 1-4 contain the values of column density ($N(H_{2})$), clouds' effective depth ($\Delta^{\prime}$), mass density ($\rho$), and velocity dispersion ($\sigma_{v}$). Rows 5-7 summarize the results from the dispersion analysis. Rows 8 and 9 correspond to values associated to the background flow. Rows 10-12 display the POS magnetic field strengths.}
\noindent $^{\rm a}$ Calculated over spatial scale equal to $\Delta^{\prime}$.
\label{tab:CND_res}
\end{table}

Because $N(H_{2})$, $\rho$, and $\sigma_{v}$ vary less than a factor of two over each streamer (Figure \ref{fig:aux_data}), we display their median in rows 1 – 4 of Table \ref{tab:CND_res}. Rows 5 -7 show the results of the dispersion analysis. Rows 8 and 9 show the lower and higher values of the large-scale flow and its Laplacian. Rows 10 - 12 show values of $B_{\rm POS}$ calculated according to the procedure described in Appendix \ref{sec:unc} -- for each DCF approximation: Classical (10), large-scale flow (11), and shear-flow (12). For each approximation, the median value is displayed along with the range corresponding to the 5 and 95 percentiles, in parenthesis.

For the full CND, the classical DCF method produces POS B-field strengths between 5.4$\pm$0.4 and 9.9$\pm$0.8 mG, with a median of 7.3$\pm$0.6 mG. The bimodal shape of the distribution in Figure \ref{fig:Bpos_hist}-Left may reflect the fact that all three streamers (with different physical conditions) are considered in one analysis. When using the large-scale modification of the DCF, $B_{\rm POS}^{\rm DCF,F}$, we obtain a range of values 0.3$\pm$1.4 - 17.3$\pm$3.2 mG, with a median of 2.9$\pm$0.9 mG. This larger range of $B_{\rm POS}$ values (in comparison to that of classical DCF) reflects the large range of flow speed present in the streamers. Finally, using the shear-flow modified DCF expression, we get lower values of $B_{\rm POS}$, between 0.3$\pm$0.5 and 2.9$\pm$0.8 mG, with a median of 1.0$\pm$0.2 mG. Magnetic fields of few milligauss ($\lesssim$ 5 mG) have been previously found in the CND \citep{Hsieh2018,Dowell2019,Schmelz2020}. However, we note that stronger B-field strengths at scales of $\sim$1 pc seem possible but not likely, based on the small occurrence of such values in our results.  

Applying the modified DCF, $B_{\rm POS}^{\rm DCF,F}$, approximation to the entire CND region is not the best approach since it assumes a B-field structure primarily in the same direction as the large-scale flow. In addition, we show that a median B-field of the CND may not be appropriate as there are different physical environments (i.e., streamers) within the considered field of view that require independent analysis.

We now perform similar calculations for each streamer separately using the masks constructed based on the POS velocity field (Figure \ref{fig:Velocity_field}). Each mask was then applied to all data sets (polarimetric, column density, and velocity dispersion) before repeating the procedure described above for the CND. All variables values and results for the three streamers are summarized in Table \ref{tab:CND_res}. First, we see that the 53-$\micron$ polarimetric data is able to resolve the turbulence scale inside the masks ($\delta > \sqrt{2}W$). It is interesting to note that the turbulence scale in the western arc, $10.25\pm0.47\arcsec = 0.38\pm0.02$ pc, is similar to that calculated for the entire CND, $11.25\pm0.27 = 0.43\pm0.01$ pc. However, in the eastern, $0.19\pm0.01$ pc, and northern, $0.12\pm0.01$, arms the turbulence scale is approximately a half and a third of the CND's value. These results reinforce the importance separate analysis for each streamer. With the results from the dispersion analysis, we computed the same estimations as those in Figure \ref{fig:Bpos_hist} for all three streamers. Using the traditional DCF approximation, median values of $B_{\rm POS}^{\rm DCF}$ are: $12.9\pm1.1$, $33.3\pm5.9$, $26.9\pm2.2$ mG for the eastern arm, northern arm, and western arc, correspondingly. All three of these values are considerably larger than the value calculated for the CND with the same classical DCF approximation. The range of values of these medians appears to be symmetric with a width of only few mG (hence, the relatively small fractional uncertainties). Although magnetic field strengths of tens of milli Gauss is likely in smaller regions in the CND, we know from the results with synthetic data that values calculated with $B^{\rm DCF}_{\rm POS}$ are overestimated by $\gtrsim 40$\%. On the other hand, values of $B^{\rm DCF,F}_{\rm POS}$ show distributions with long tails at large values and no apparent trend for the median. As mentioned before, this is due to the break down of the requirement that $\sigma_{\phi}(U_{0}/\sigma_{v})>1$ for the $B^{\rm DCF,F}_{\rm POS}$ approximation. Thus, the $B_{\rm POS}^{\rm DCF}$ and $B_{\rm POS}^{\rm DCF,F}$ are not considered in the scientific analysis of the CND that follows.

Finally, using the shear-flow modified DCF ($B^{\rm DCF,SF}_{\rm POS}$) we obtain physically reasonable median values for all three streamers. We estimate $4.0\pm1.2$ mG, $4.7\pm1.3$ mG, and $8.5\pm2.3$ mG for the eastern arm, northern arm, and western arc, respectively. These results suggest that both northern and eastern arms have similar median POS magnetic field strengths but less-common values greater than 10 mG (based on the 95th percentile of their $B_{\rm POS}^{\rm DCF,SF}$ distributions). The western arc, on the other hand, shows stronger magnetic fields than those in the other two arms -- approximately two times their values. These larger values in the western arc can be attributed to the fact that this particular streamer displays the lowest range of shear velocity and its Laplacian, which is expected according to Eq. \ref{eq:mod_dcf}.

\section{Discussion}\label{sec:discussion}

\subsection{B-fields in the gas streamers of the CND}\label{subsec:B-field_gastreamer}

The B-field morphology is largely organized and cospatial with the ionized gas streamers pointing inwards toward SgrA*. Our modified DCF method accounting for the shear flow suggests that these streamers have POS B-field strengths of $4-9$ mG with typical turbulent length scales of $0.1-0.4$ pc ($3.2\arcsec-10\arcsec$). We discuss these results in terms of previously reported values of the B-field strength and the dynamics of the CND.

Zeeman measurements have inferred B-field strengths of $1-3$ mG (pointing to us) in the LOS direction in the southern and northern regions of the CND \citep{Killeen1992,Plante1995}. These results were obtained with OH and HI gas with a beam size of $4\arcsec-6\arcsec$ using the VLA. Preliminary results using the classical DCF method with 53-$\mu$m HAWC+/SOFIA data reported similar values for the POS B-field component \citep{Dowell2019}. However, as stated previously, these estimates are subject to the limitations of the classical DCF approximation, specially in presence of large-scale and sheared flows. Thus, the B-fields in the CND need to be characterized for each streamer.

The western arc has a larger $B_{\rm POS}$ value than the other two streamers in the CND. Given that the shear-flow speeds in the western arc are lower than those in the northern and eastern arms, one might speculate that the western arc has not been disrupted as much by the gravitational pull of Sgr A*. In addition, this streamer has the smallest width and the highest column density. We conclude that its B-field strength may be closer to that of the unperturbed B-field initially at equilibrium with the gas and dust and/or highly compressed within the streamer.

\begin{figure}[h!]
    \centering
    \includegraphics[width=0.6\textwidth]{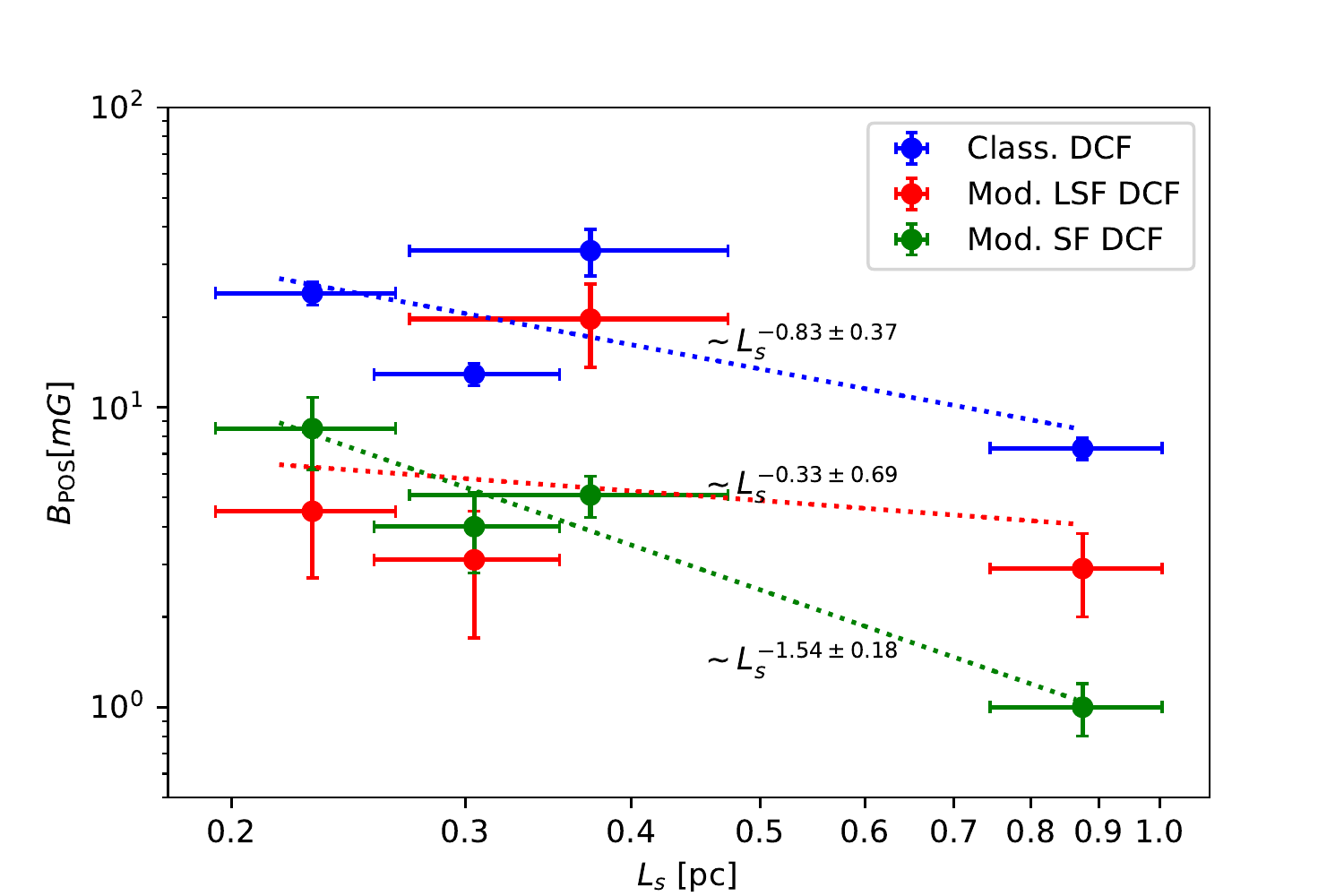}
    \caption{POS magnetic field strength as a function of spatial scale size, $L$. Three values are shown: blue symbols correspond to $B^{\rm DCF}_{\rm POS}$; red symbols correspond to $B^{\rm DCF,F}_{\rm POS}$ while green symbols correspond to $B^{\rm DCF,SF}_{\rm POS}$. In all three cases, $B_{\rm POS}$ scales as a power law with $L_{s}$.}
    \label{fig:bpos_scale}
\end{figure}

On the other hand, we can see from Table \ref{tab:CND_res} that the $B_{\rm POS}$ values seem to scale with the ``size" of the cloud: larger areas tend to have weaker magnetic fields. In order to characterize this tendency, Figure \ref{fig:bpos_scale} shows $B_{\rm POS}$ as a function of the one-dimensional length scale, $L_{s}$. Here, assuming that the distribution of the gas is isotropic, we can use the values of $\Delta^{\prime}$ to estimate $L_{s}$. We display all four median values (the three streamers and the CND from Table \ref{tab:CND_res}) for the classical DCF approximation (blue), the large-scale flow modified DCF (red), and the shear-flow modified DCF (green). The classical DCF and shear-flow modified DCF are well described by power-law functions, but the large-scale flow modified DCF is not. For those cases well-described by power laws, the best fits have exponents of -0.83$\pm$0.37 and -1.54$\pm$0.18, respectively, which are larger than the value suggested by \citet{Chuss2003} of -0.5. The steeper dependence of $B_{\rm POS}$ with $L_{s}$ results because the polarization angle dispersion -- not explicitly considered in the \citet{Chuss2003} analysis -- also depend on $L_{s}$, which can appear in non-linear terms when large-scale flows are considered. Therefore, this might indicate the existence of stronger magnetic fields in smaller regions when shear flows interact with it---i.e., B-field amplification occurs to conserve the magnetic flux per area.

Based on our results, we can hypothesize that the high-speed velocity field in the streamers might affect turbulent and ordered magnetic fields in different ways. Following \citet{Dinh2021} that concluded from smooth particle hydrodynamics that streamers in CNDs can be transient features caused by large-scale ($0.03 - 2$ pc) turbulence with lifetimes of $\sim$10$^{5}$ years, we can assume that the large-scale ordered B-field evolves at a temporal scale much larger than that of the turbulent motions. Therefore, the presence of the shear flow in the streamers may provide a restoring force that acts against the turbulent motions perpendicular to the flow-magnetic field direction. This appears evident in the northern arm and western arc: their velocity dispersion values (proxy for turbulent motions velocity) are similar within the uncertainties, but the northern arm displays shear speeds and lower values of angular dispersion in comparison with those in the western arc. This does not appear to be the case for the eastern arm, but in this case the velocity dispersion is larger than in the other two streamers. As a consequence, even relatively weak magnetic fields can display low angular dispersion.

On the other hand, the effect of the shear flow on the large-scale, ordered magnetic field can be more complex. Although $B^{\rm DCF,SF}_{\rm POS}$ does not depend explicitly on spatial coordinates, it does through the spatial dependence of the shear flow. It is clear from Figure \ref{fig:Velocity_field} that locations with slow flow have larger radial distances from the orbit's focal point (Sgr A*). For shorter radial distances, the flow has larger speeds. Through Eq. \ref{eq:mod_dcf} one can expect to find weaker magnetic fields where the flow is faster. This dependence of the magnetic field strength with radial distance poses an important aspect for the energetics in the CND and the streamers. This is studied in the following section (see Figure \ref{fig:CND_bpos_beta}).

\subsection{Importance of the B-field in the CND}\label{subsec:plasma_beta}

To assess the importance of the B-field in comparison to the thermal gas dynamics, we measure the plasma $\beta$ parameter -- the ratio of gas thermal pressure to magnetic field pressure. Assuming a conservative value for gas temperature $T_{\rm gas}=$ 300 K \citep{Mills2013} and values from Table \ref{tab:CND_res} we calculate the plasma $\beta$ parameter as 

\begin{equation}
    \beta_{\rm plasma} = \frac{n(H_{2})k_{B}T_{\rm gas}}{\left( B^{\rm DCF,SF}_{\rm POS} \right)^{2}/8\pi}.
\end{equation}

\noindent
where $n(H_{2}) = N(H_{2})/\Delta^{\prime}$ is the number density. We found that the range of values of this parameter for the streamers are: $\beta_{\rm plasma}=$1.6$\times$10$^{-3}$ - 2.3$\times$10$^{-2}$ for the northern arm, $\beta_{\rm plasma}=$ 2.7$\times$10$^{-4}$ - 5.9$\times$10$^{-2}$ for the eastern arm, and $\beta_{\rm plasma}=$0.6$\times$10$^{-4}$ - 5.4$\times$10$^{-3}$ for the western arc. For values of $\beta_{\rm plasma}<1$ the magnetic pressure dominates, implying that the magnetic field is rigid and the gas flows along it. However, in environments where the gas kinematics has other important components, such as in the CND and streamers, a more appropriate parameter was proposed by \citet{Lopez2021} 

\begin{equation}
    \beta^{\prime} = \frac{U_{\rm H}+U_{\rm K}}{U_{\rm B}},
\end{equation}

\noindent
which compares the sum of hydrostatic ($U_{\rm H}$) and turbulent kinetic ($U_{\rm K}$) energy densities to that of the magnetic field ($U_{\rm B}$). We use the following definitions $U_{\rm H} = \pi G (N(H_{2}) m_{\rm H} \mu )^{2}$, $U_{\rm K} = \frac{1}{2}\rho\sigma_{v}^{2}$, and $U_{\rm B} = \left(B^{\rm DCF,SF}_{\rm POS}\right)^{2}/8\pi$, where $N(H_{2}) m_{\rm H} \mu$ is the gas area density with $m_{\rm H}=$1.67$\times$10$^{-24}$g being the mass of hydrogen, $\mu=2.8$ the mean molecular weight, and $G$ is the gravitational constant. The range is $\beta^{\prime}=$7.7$\times$10$^{-2}$ - 9.4$\times$10$^{0}$ for the northern arm, 
$\beta^{\prime}=$1.6$\times$10$^{-1}$ - 3.0$\times$10$^{2}$ for the eastern arm, and 
$\beta^{\prime}=$5.8$\times$10$^{-2}$ - 1.2$\times$10$^{1}$ for the western arc.

\begin{figure}[h!]
    \centering
    \includegraphics[width=0.7\textwidth]{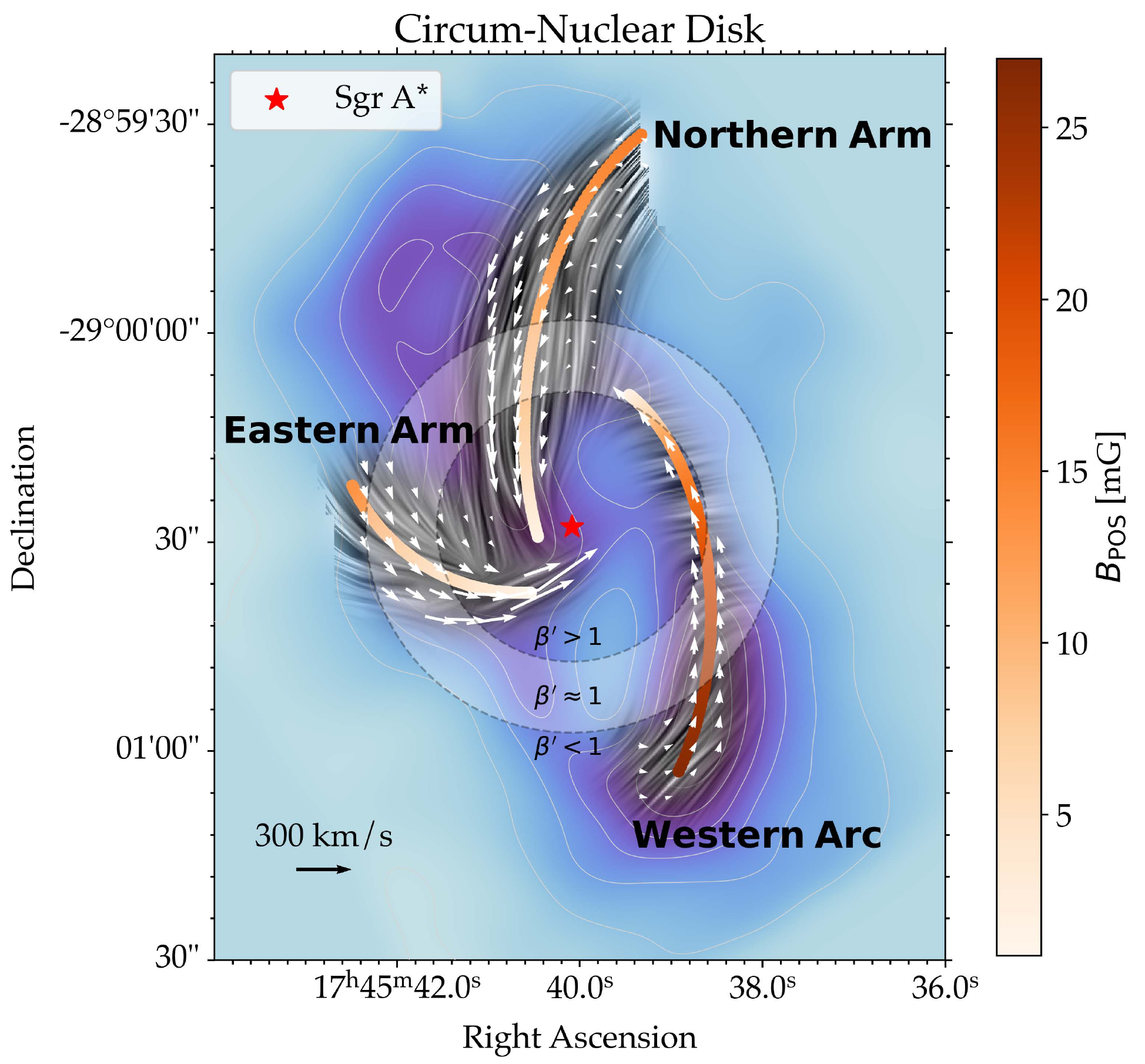}
    
    \caption{B-field configuration in the CND and streamers. B-field's orientation is represented for each streamers by line integration contours \citep[LIC;][]{Cabral1993} inferred from the 53-$\micron$ HAWC+ data. B-field's strength is represented in a one-dimensional way along the main direction of the streamer. Stronger B-fields correspond to larger distances away from Sgr A* (red star) and weaker magnetic fields are inferred closer to Sgr A* where the shear flows (white arrows) are larger. Background image corresponds to the 53-$\micron$ intensity. A legend showing a velocity field of $300$ km s$^{-1}$ is displayed.}
    \label{fig:CND_bpos_beta}
\end{figure}

Remembering that the inferred B-fields are stronger where the streamer flows are slower (larger radial distances from Sgr A*) and weaker where the flows are faster (smaller radial distances), the $\beta^{\prime}$ results imply that a transition from magnetically-dominated ($\beta^{\prime}<1$) to gravitationally-dominated ($\beta^{\prime}>1$) accretion of CND material onto Sgr A* occurs somewhere along the streamer's long axis. This transition occurs where $\beta^{\prime}\approx1$. In order to visualize this transition, Figure \ref{fig:CND_bpos_beta} displays a 1-D representation of the magnetic field dependence on the CND, atop the inferred magnetic field orientation. For each streamer, the color bar represents the variation of the magnetic field along the shear flow main direction. Stronger fields are at the outskirts of the streamer while weaker fields are at the inner regions; {\it i.e.,} closer to Sgr A*. Using the values calculated in this work, we determine that the transition takes place at projected radial distances from Sgr A* between 0.7 pc and 1.8 (mean: 1.1 pc) for the northern arm, between 0.5 and 1.3 pc (mean: 0.8 pc) for the eastern arm, and between  1.1 and 1.3 pc (mean: 1.2 pc) for the western arc. The variation of these values occur primarily in the direction perpendicular to the streamer's axis, with the smaller value in the outer edge of the POS velocity field. The transition area ($\beta \approx 1$) is displayed in Figure \ref{fig:CND_bpos_beta} as a shaded annulus, with inner and outer radii defined by the maximum and minimum values of the means above, 0.8 and 1.3 pc, correspondingly.

\citet{Hsieh2021} found, using ALMA CS observations, that turbulence in the CND changes the limits of densities in the vicinity of Sgr A* for which dense clouds can collapse and form stars. For distances $\geqslant$1.5 pc turbulence dominates the internal energy and clouds require smaller densities to collapse. For distances $\leqslant$ 1 pc, tidal forces dominates and the density limit quickly increases. Our results also demonstrate that at distances $\lesssim$ 1 pc from Sgr A*, gravitational pull is very strong and dominates over magnetic forces as well. For distances $\gtrsim$ 1 pc, on the other hand, magnetic fields -- not considered in \citet{Hsieh2021} -- can affect such collapsing density limits by the well-known fact that collapsing perpendicular to the magnetic field lines is prevented by increased magnetic pressure.

In regimes of large to very large magnetic Reynold's number, $R_{m}$, like in the CND\footnote{$R_{m} \approx 10^{3} -10^{23}$ for the warm and hot gas in the discs of spiral galaxies (\citet{shukurov_subramanian_2021}, Table 2.2).}, the convection of the magnetic field dominates over the diffusion in the evolution of the magnetic field. Because of the shearing nature of the steady-state flow in the streamers of the CND, the magnetic field weakens as the flow speeds up and become more sheared. This picture appears to be consistent with our observation that the magnetic field along the direction of the streamers is weaker closer to Sgr A*. Moreover, because of this transition from magnetically-dominated to gas-dominated dynamics in the streamers, one can conclude that for radii $\gtrsim$ 1 pc (on average) from the Sgr A* the accretion of material might be influenced by the magnetic field. Determining the nature of such influence will likely require a detailed knowledge of the B-field strength (map) obtained by higher resolution observations or numerical MHD simulations of the CND.

\section{Conclusions}\label{sec:summary}

We have studied the effect of shear flows on the strength of the POS magnetic field in the CND using dust polarimetric observations from SOFIA/HAWC+. This study required modifying the DCF method by studying the propagation of an Alfv\'en wave along a magnetic field with a background, structured flow. As in the case of the classical DCF, the wave dispersion relation provides the means to infer the magnetic field. We obtained an expression for the POS magnetic field strength, which we called the shear-flow modified DCF, that depends explicitly on the spatial structure of the flow and its Laplacian. This new DCF approximation reduces to the classical DCF when no background flow term is present, and the large-scale expression developed by \citet{Lopez2021} is recovered for a constant flow. Moreover, when the amplitude of the shear flow is much larger than the velocity dispersion, the modified DCF expression tends to behave similarly to that of \citet{Skalidis2021}.

This modified DCF approximation was first tested using synthetic polarimetric data constructed from MHD simulations. These simulations were performed by the CATS project and made available for public use through their website. We chose a simulation with physical conditions that resembled those of the CND: sub-sonic, super-Alfv\'enic turbulence in a hot, low$-\beta$ plasma. This test revealed that the modified DCF expression provided values of $B_{\rm POS}$ with errors between 3\% and 22\% and fractional uncertainties between 5\% and 13\%, for lines-of-sight perpendicular to the initial uniform magnetic field. A big contribution to the uncertainties and errors comes from the terms related to the background flow structure.

The magnetic field strength in the CND was estimated using the modified DCF expression. We used SOFIA/HAWC+ 53-\micron\ polarimetric data to estimate the dispersion in the magnetic field and column densities from SED fittings. Velocity dispersion values were determined from spectral fitting of [CII] observations taken with SOFIA/GREAT. The POS velocity field of the sheared flow in the CND streamers was determined by Keplerian orbital modeling. We found that the magnetic field strength for the entire CND ranges between $\sim$0.3 - 3 mG, which is not uncommon at these spatial scales. Stronger magnetic fields were found in all three streamers ($\sim$1 - 14 mG for the northern arm; $\sim$1 - 16 mG for the eastern arm; $\sim$3 - 27 mG for the and western arc). Stronger magnetic fields correspond to locations in the streamers where the flow is slower, further away from Sgr A*. Weaker magnetic fields, in turn, are the result of faster shear flows, typically occurring closer to Sgr A*. In addition, we found that the dependence of $B_{\rm POS}$ with the length scale of the cloud follows an inverse power law, with exponents larger than previously reported \citep{Chuss2003}.

The presence of the steady-state, shear flow appears to have two separate effects on the components of the magnetic field: 1) the flow along the magnetic field direction can reduce the turbulent component by suppressing motions perpendicular to the magnetic field; 2) although the magnetic field pressure seems to dominate over the thermal gas motions, other important kinematic properties (i.e., turbulence) in the streamers provide conditions where the energy density in the gas seems to dominates throughout most of the CND. Our results also suggest that the magnetic field is weakened along the direction of the flow (towards Sgr A*) due to its shearing nature. Finally, our results indicate that the transition from magnetically to gravitationally dominated accretion of material onto Sgr A* occurs at distances $\gtrsim$ 1 pc.

More accurate values of POS magnetic field strength require measuring the dependence of the angular dispersion ($\sigma_{\phi}$) along and perpendicular to the magnetic-field-flow structure, which can only be achieved by higher resolution polarimetric data from observatories like ALMA. In addition, velocity dispersion values can be improved by obtaining observations of molecular tracers coexisting with the ionized material in the streamers (rather than the dust). In future we want to improve upon this work by better modelling the three-dimensional geometry of the streamers in order to determine more accurate line-of-sight cloud depths, which in turn improves parameters from the dispersion analysis and mass density estimates.

\begin{acknowledgments}
Based on observations made with the NASA/DLR Stratospheric Observatory for Infrared Astronomy (SOFIA). SOFIA is jointly operated by the Universities Space Research Association, Inc. (USRA), under NASA contract NAS2-97001, and the Deutsches SOFIA Institut (DSI) under DLR contract 50 OK 0901 to the University of Stuttgart. Financial support for this work was provided by NASA through awards \#SOF 09\_0194 and \#SOF 09\_0537 issued by USRA. The authors thank the anonymous reviewer, whose comments helped improve this paper.
\end{acknowledgments}

%

\vspace{5mm}
\facilities{SOFIA/HAWC+, SOFIA/GREAT}


\software{ \texttt{python, Ipython} \citep{Perez2007}, \texttt{numpy} \citep{vanderWalt2011}, \texttt{scipy} \citep{Jones2001} \texttt{matplotlib} \citep{Hunter2007}, \texttt{emcee} \citep{Foreman-Mackey2013}, \texttt{corner} \citep{Foreman-Mackey2016}, \texttt{astropy} \citep{2018AJ....156..123A, 2013A&A...558A..33A}, \texttt{yt} \citep{Turk2011T}}, \texttt{Uncertainties}\citep{LebigotUnc}.



\appendix

\section{Dispersion Analysis}\label{sec:app_a}

In the DCF approximation (Eqs. \ref{eq:DCF_og}, \ref{eq:DCF_lsflow}, \ref{eq:mod_dcf}), the measure of angle dispersion $\sigma_{\phi}$ should only consider the deviations of the magnetic field due to the turbulent motions and not deviations due to the overall shape of the magnetic field. \citet{Hildebrand2009,Houde2009} developed a method (which we call {\it dispersion analysis} throughout this paper) that measures $\sigma_{\phi}$ by characterizing the structure function of the angle difference $\Delta\phi(\ell)$ between all pair of angles separated by the distance $\ell$, $\dispfunct$. This function is then described by a two-scale model,

\begin{equation}
    \dispfunct \approx \frac{1-e^{ -\ell^{2}/2(\delta^{2} + 2W^{2})}}{1 + \mathcal{N}\left[\frac{\langle B_{t}^{2}\rangle}{\langle B_{0}^{2}\rangle}\right]^{-1}} + a_{2}\ell^{2}.
    \label{eq:disp_funct}
\end{equation}

\noindent
where the first term describes the contribution to the dispersion from the small-scale, turbulent motions, and the second term describes the contribution from the large-scale magnetic field geometry. In Eq. \ref{eq:disp_funct} $\delta$, $W$, $\mathcal{N}$, and $\ratioinline$ correspond to the turbulence correlation length, the observation's beam radius, the number of turbulent cells along the line-of-sight defined as $\mathcal{N}=(\delta^{2}+2W^{2})\Delta^{\prime}/(\sqrt{2\pi}\delta^{3})$, and the ratio of turbulent-to-ordered magnetic field energies, respectively. The first term is the most relevant for the DCF estimation because it considers the correction to the dispersion due to the line-of-sight (LOS) and (assumed Gaussian) beam integration. In this context, $\sigma^{2}_{\phi}\approx b^{2}(0)$, where $b^{2}(0)=(\ratioeq)\mathcal{N}^{-1}$ is the amplitude of the correlated turbulent part of the dispersion function (first term in Eq.\ref{eq:disp_funct}). Thus, the angle dispersion can be estimated as $\sigma_{\phi}=\sqrt{\mathcal{N}b^{2}(0)}=\sqrt{\ratioeq}$.

\begin{figure}[h!]
    \centering
    \includegraphics[width=0.5\textwidth]{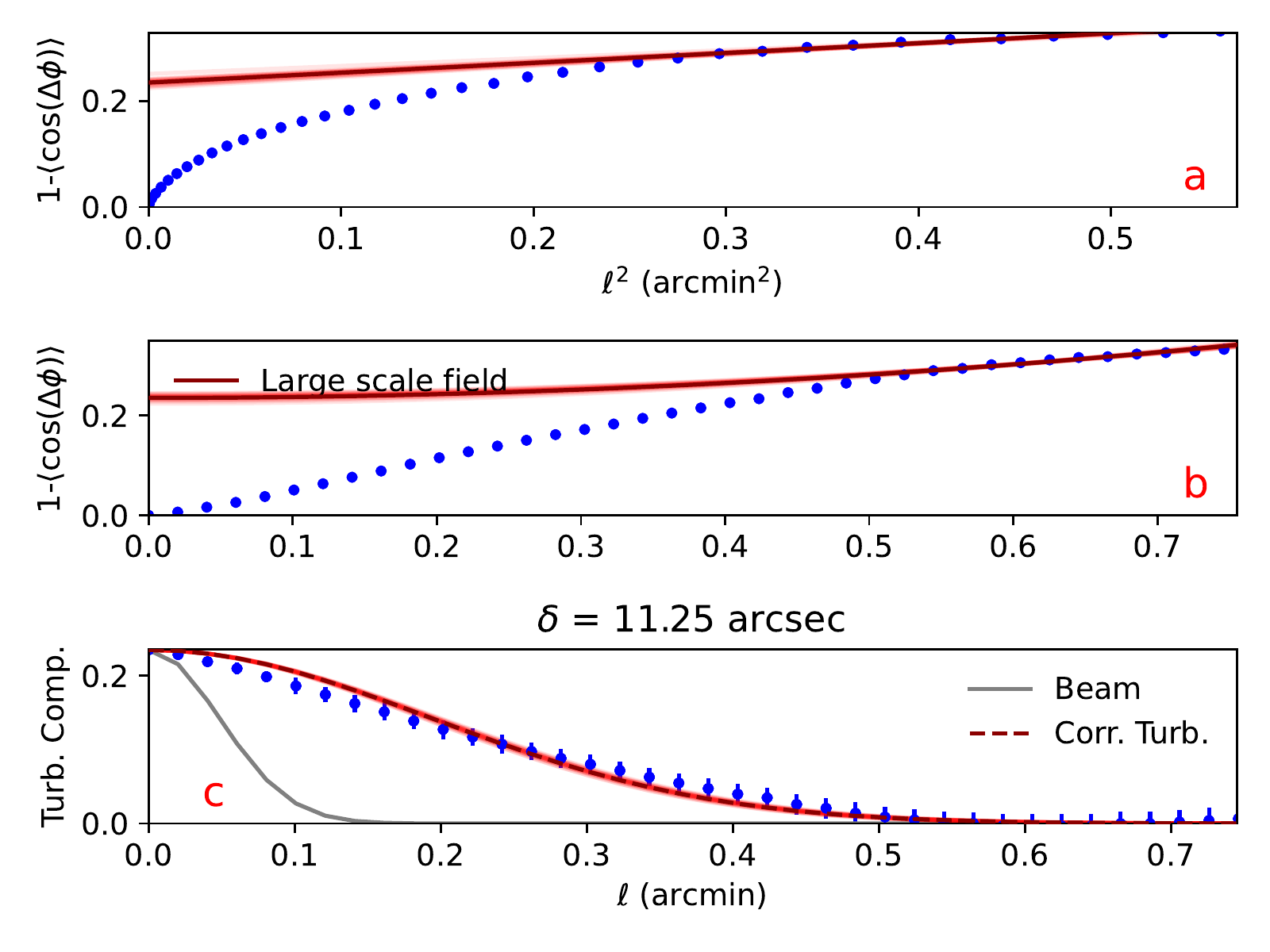}
    \includegraphics[width=0.4\textwidth]{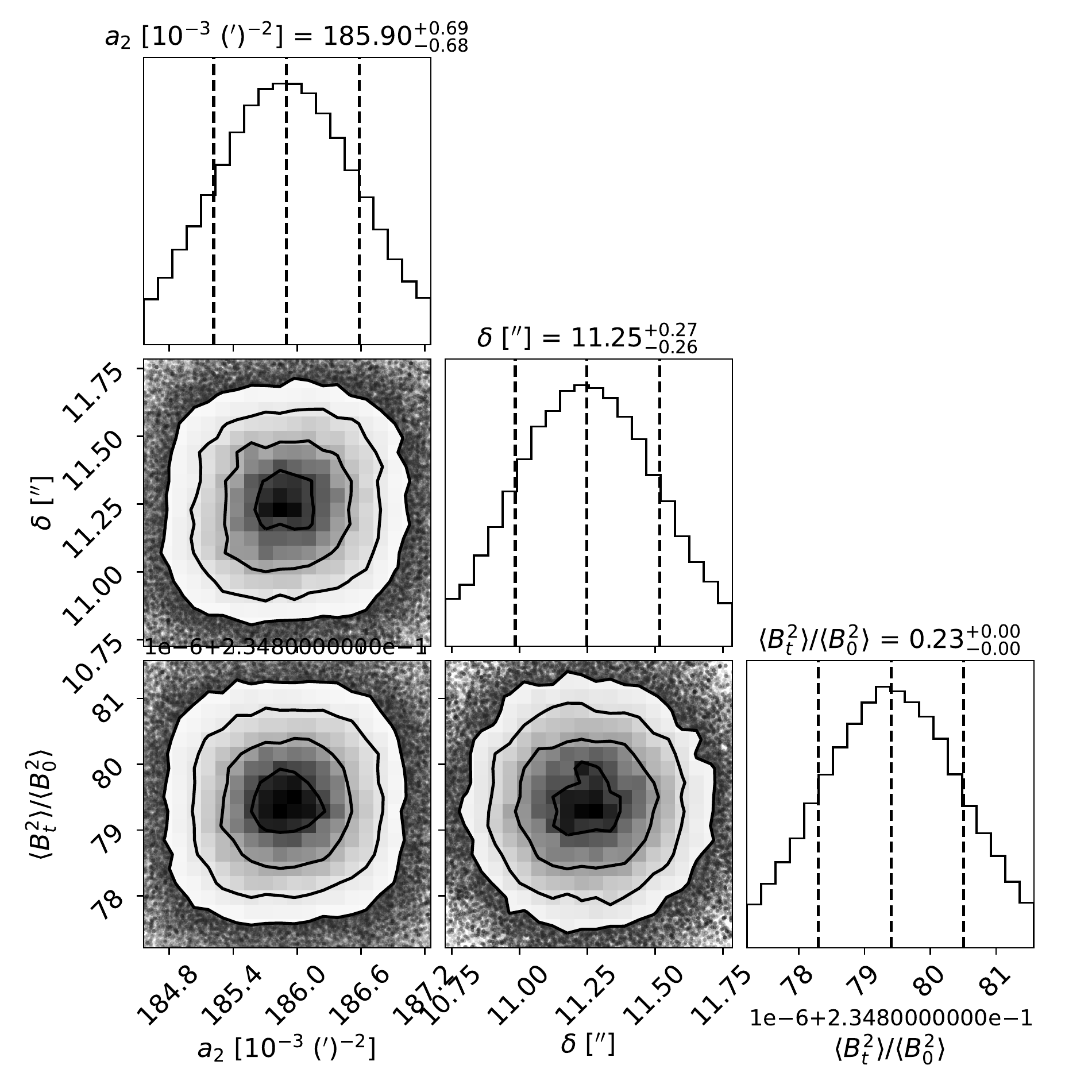}
    \caption{Results from the dispersion analysis and corresponding MCMC fitting performed to the entire CND. {\it Left:} Dispersion function (blue circles) as a function of $\ell^{2}$(panel a) and $\ell$ (panel b). Solid red line corresponds to the best-fit model. Panel c displays the auto-correlated turbulent component of the dispersion function (blue circles) with its corresponding fit (red dashed line). For comparison, the auto-correlated beam is displayed with the solid grey line. {\it Right:} Corner plots for the MCMC posterior distributions of parameters fitting the dispersion function.}
    \label{fig:disp_analysis_ex}
\end{figure}

Parameters $a_{2}$, $\delta$, and $\ratioeq$ (or $b^{2}(0)$) are determined from dispersion functions like those in Figure \ref{fig:disp_analysis_ex}{\it Left}. In this Figure, three panels are shown: a) the dispersion function (blue circles) where the best-fit model (red line) for the large scale component is displayed as a function of $\ell^{2}$; b) same as a) but as a function of $\ell$; c) the correlated turbulent component (blue circles), $\dispfunct - a_{2}\ell^{2}$, the corresponding best fit (red dashed line), and the correlated beam profile (grey solid line). The importance of panel c is to verify that  the correlated turbulent function is wider than the uncorrelated beam ({\it i.e.,} $\delta > \sqrt{2}W$), meaning that the turbulence is resolved by the polarimetric observations and the effect of beam-and-LOS integration is properly accounted for in the angular dispersion estimation.

Throughout this work, the best-fit parameters  for the dispersion functions were determined using a Markov-Chain Monte-Carlo (MCMC) algorithm, \texttt{emcee}. In Figure \ref{fig:disp_analysis_ex}({\it Right}), corner plots show the posterior distribution and correlation plots for parameters $a_{2}$, $\delta$, and $b^{2}(0)$. Best-fit parameter values are estimated as the median values and their uncertainties correspond to the standard deviation.

Figure \ref{fig:disp_analysis_ex} specifically displays the results of the dispersion analysis for the entire CND. In panel c({\it Left}), it is clear that the turbulence in the gas and magnetic field is resolved by the observation's beam size, since the correlated turbulent function is wider than the autocorrelated beam profile (or $\delta = 11.25^{\prime\prime} > \sqrt{2}W = 2.90^{\prime\prime}$). Therefore, the value of $\sigma_{\phi}$ can be regarded as accurate and the resulting $B_{\rm POS}$ values are not overestimated. Similar analyses were performed for the CND streamers separately (not shown here). All resulting values are summarized in Table \ref{tab:CND_res}, rows 5 - 7.

\section{Velocity Dispersion Data}\label{sec:app_b}

\begin{figure}[h!]
    \centering
    \includegraphics[width=0.45\textwidth]{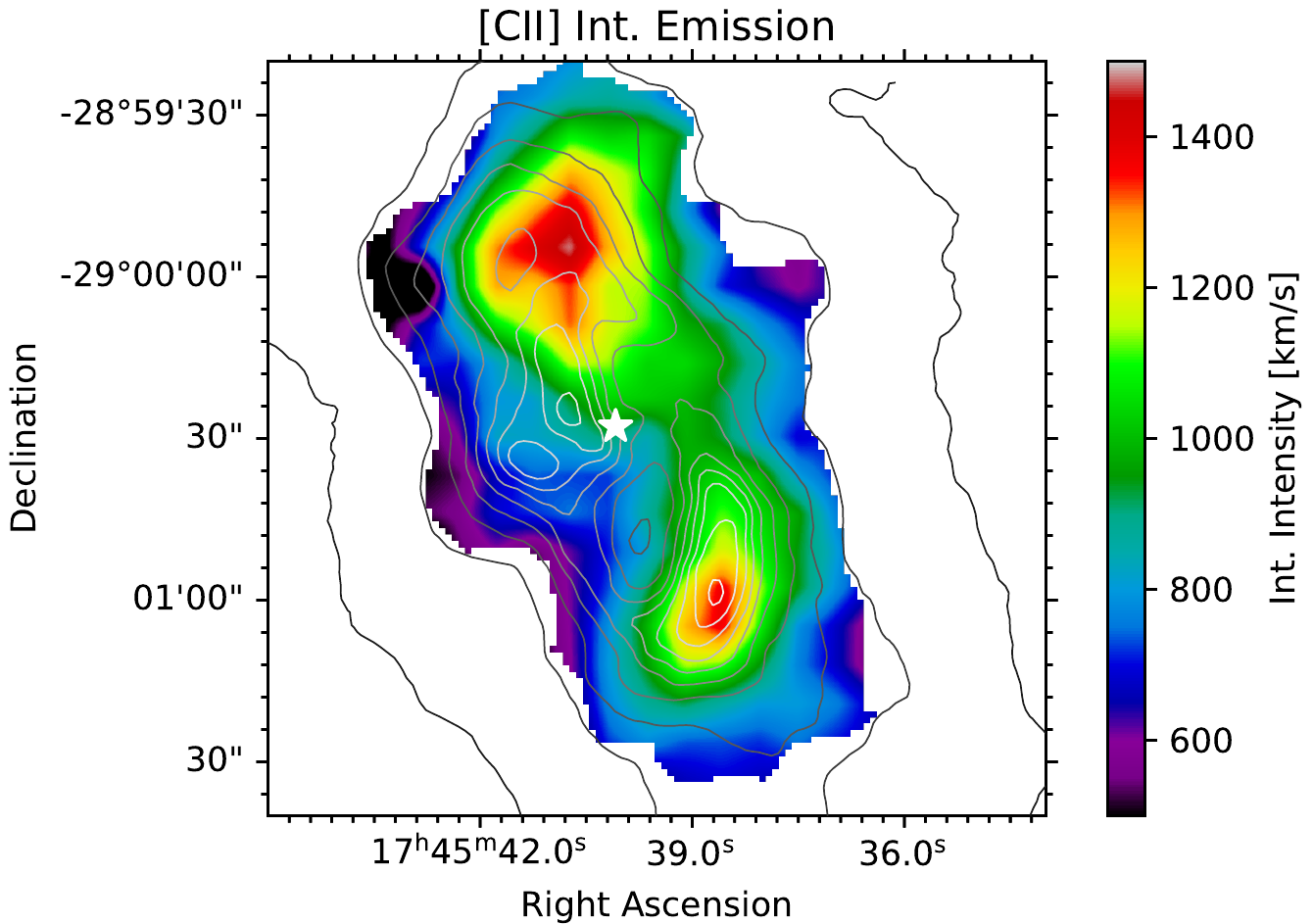}
    \includegraphics[width=0.45\textwidth]{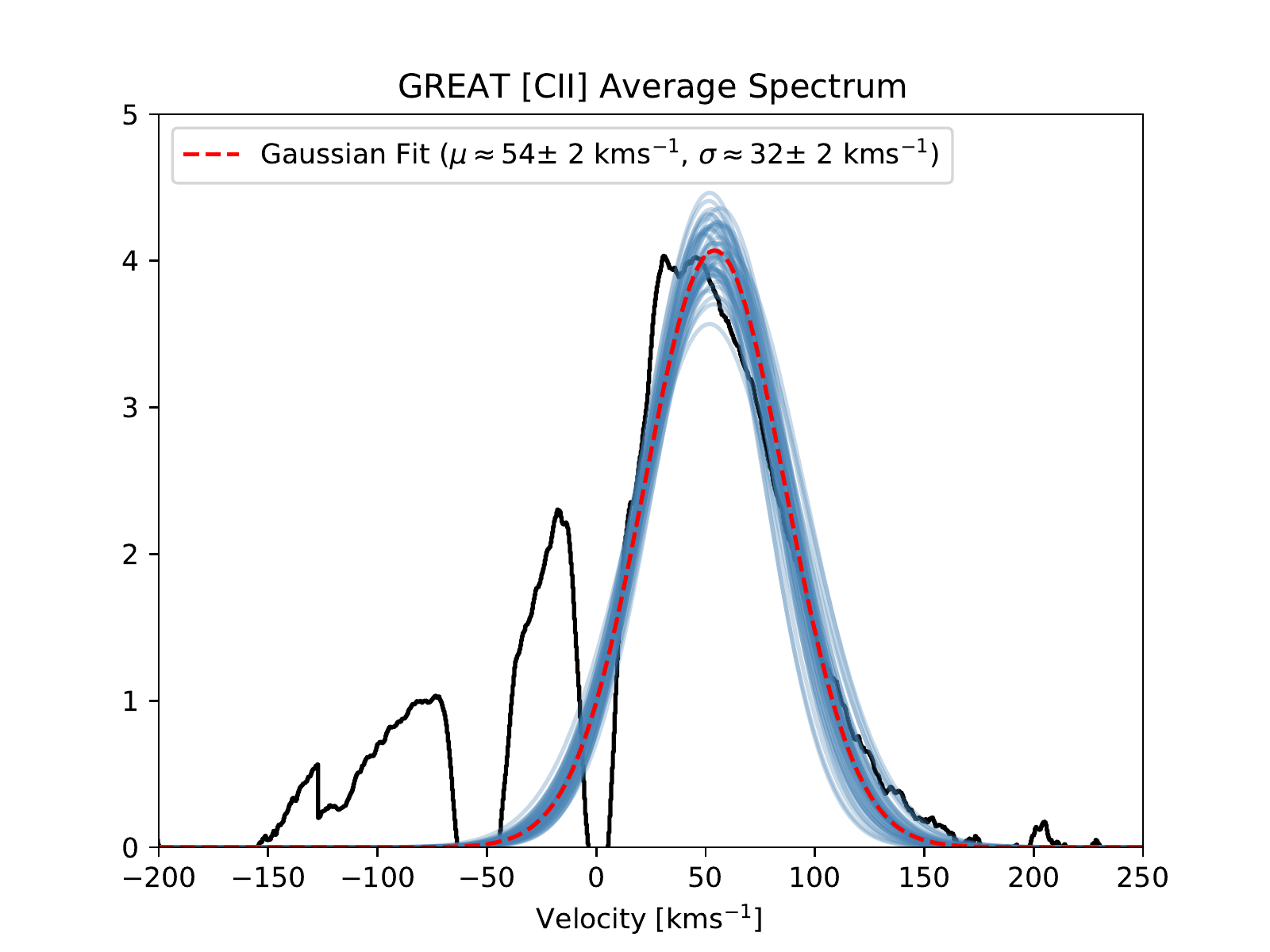}
    \caption{Average spectrum of [CII] emission from the CND. Observations were obtained with SOFIA/GREAT. The emission line was modelled as Gaussian in order to estimate its width as the turbulent velocity dispersion. Measured line widths were corrected for thermal motions widening. Figure \ref{fig:aux_data}({\it Right}) displays the map of velocity dispersion derive from this data.}
    \label{fig:CII_ave_spec}
\end{figure}

In order to determine the velocity dispersion map for the CND, we used [CII] observations of the CND by SOFIA/GREAT (cycle 5, proposal 05\_0102, PI: Morris, M.). We used the \texttt{python} package \texttt{spectral\_cube} to manipulate the data cube and perform the reduction. The following reduction steps were taken to measure the line widths: i) data were smoothed in the spectral direction was using a 1D box kernel (window) with a size of 20 velocity steps\footnote{Variations of this kernel's size resulted in little to no effect in line widths.} (7.7 \kms); ii) a spectral noise map was calculated by determining the standard deviation of the signal in the line-free region of the spectrum; iii) a spatial mask was applied to eliminate pixels with peak intensity less than three times the noise level; iv) an spectral mask was applied to each pixel to limit the spectrum to a defined width around the peak velocity; and v) maps of the moments 0-3 were created with the resulting, clean spectra. First, we determine the adequacy of the [CII] data as tracer of the turbulent motion in the CND. Figure \ref{fig:CII_ave_spec},{\it Left} displays the [CII] (spectral) integrated intensity ($M_{0}$) in the CND. Contours corresponding to the 53-\micron intensity (Figure \ref{fig:pol_data}) are overlaid on this map. We see from this figure that enhanced [CII] intensity appears to be co-spatial with the dust emission in the western arc and part of the northern arm but not with the eastern arm. This is likely due to different temperatures in the eastern arm.

Figure \ref{fig:CII_ave_spec},{\it Right} shows the spatially-averaged spectra. Two absorption signatures are seen at velocities $\sim$ -60 \kms and 0 \kms, which are known to be due to the 3 kpc spiral arm in front of the galactic center and and other clouds within few kpc of the solar neighbourhood \citep{Goto2002}. A Gaussian profile seems to fit the [CII] emission line adequately -- most emission is seen in excess of the Gaussian profile for velocity channels $<$ 0 \kms. We fitted the [CII] line using a MCMC routine so the uncertainty in the Gaussian parameters can be assessed. The dashed red line represents the best-fit Gaussian model while the blue lines are random selection of MCMC chains which represent the spread of the fitted parameters. The measured velocity dispersion values is calculated as $\sigma_{v,m}=\sqrt{M_{2}}$, where $M_{2}$ is the spectral second moment. These measured velocity dispersion values must then be corrected to extract the contribution due to thermal motions and leave only the component due to the turbulent motions \citep[see e.g.,][]{Tram2022}. Thus, we calculate $\sigma_{v}^{2} = \sigma_{v,m}^{2} - k_{B}T_{\rm gas}/m$ where $k_{B}$ is the Boltzmann constant, $T_{gas}$ is the temperature of the gas (proxy for excitation temperature), and $m$ is the ion's mass. Since [CII] is singly ionized carbon, its mass is practically same as the $C$ atom, 1.99$\times$10$^{-23} g$. CND's $T_{gas}$ map was also obtained from the  HiGAL project's public products \citep{Molinari2010}. The resulting velocity dispersion map is shown in Figure \ref{fig:aux_data}({\it Right}). Values range between $\sim$40 and $\sim$75 \kms, similar to values estimated using single-point observations of H92$\alpha$ emission \citep{Zhao2009} and H$_{2}$ \citep{Mills2017} lines, after adjusting for differences in beam size.

\section{Uncertainties in POS Magnetic Field Strength}\label{sec:unc}

\begin{figure}[h!]
    \centering
    \includegraphics[width=0.3\textwidth]{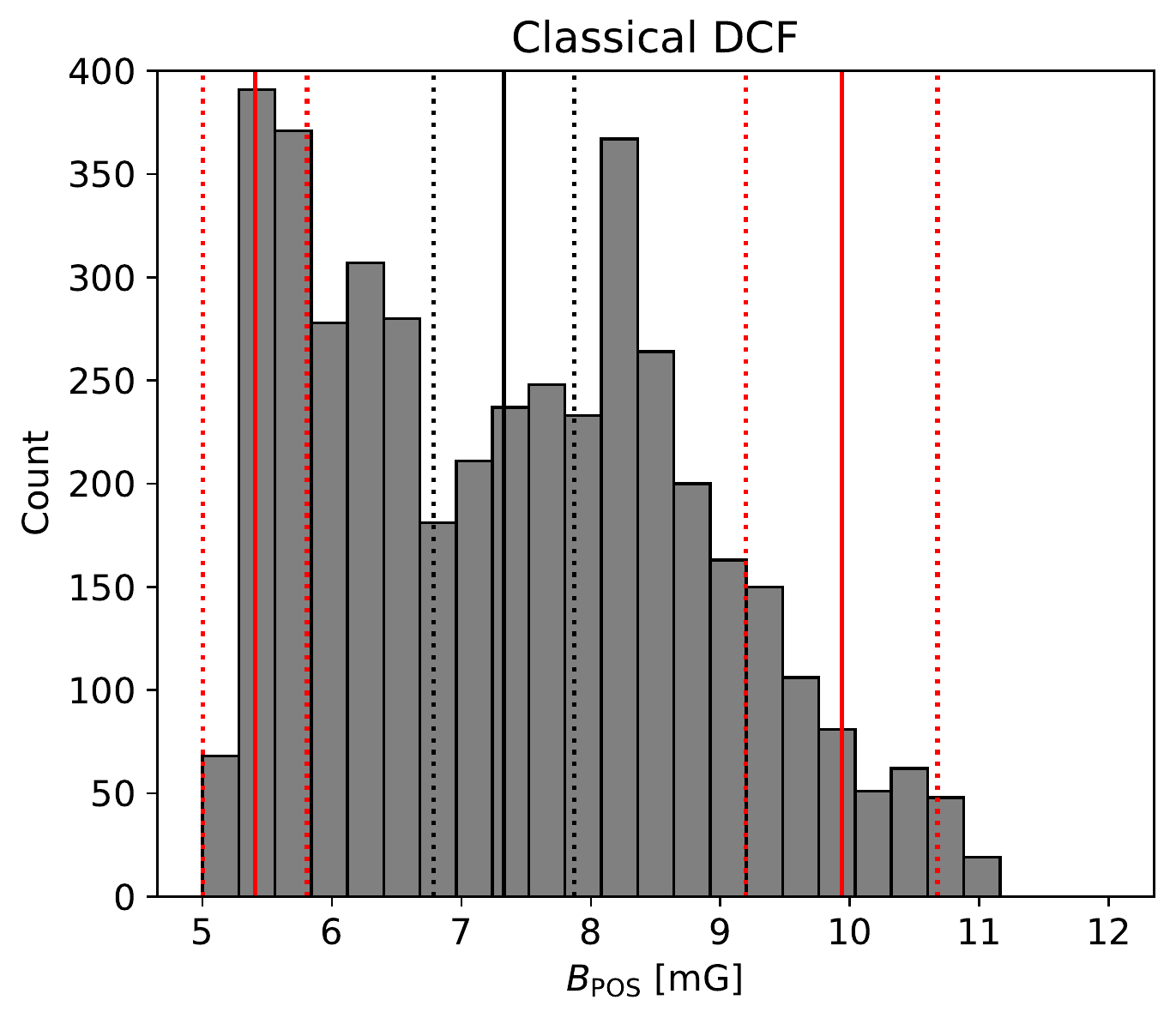}
    \includegraphics[width=0.3\textwidth]{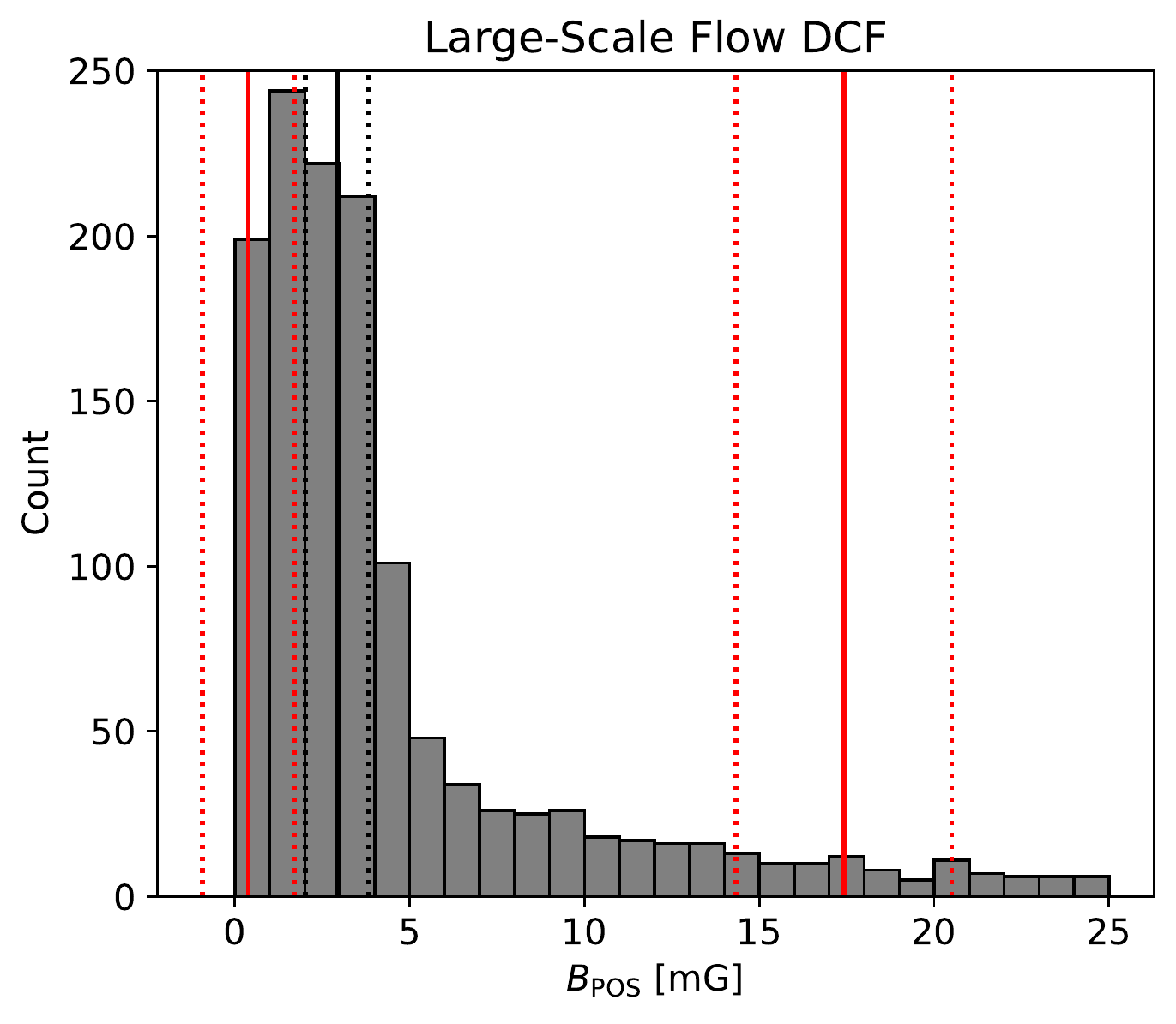}
    \includegraphics[width=0.3\textwidth]{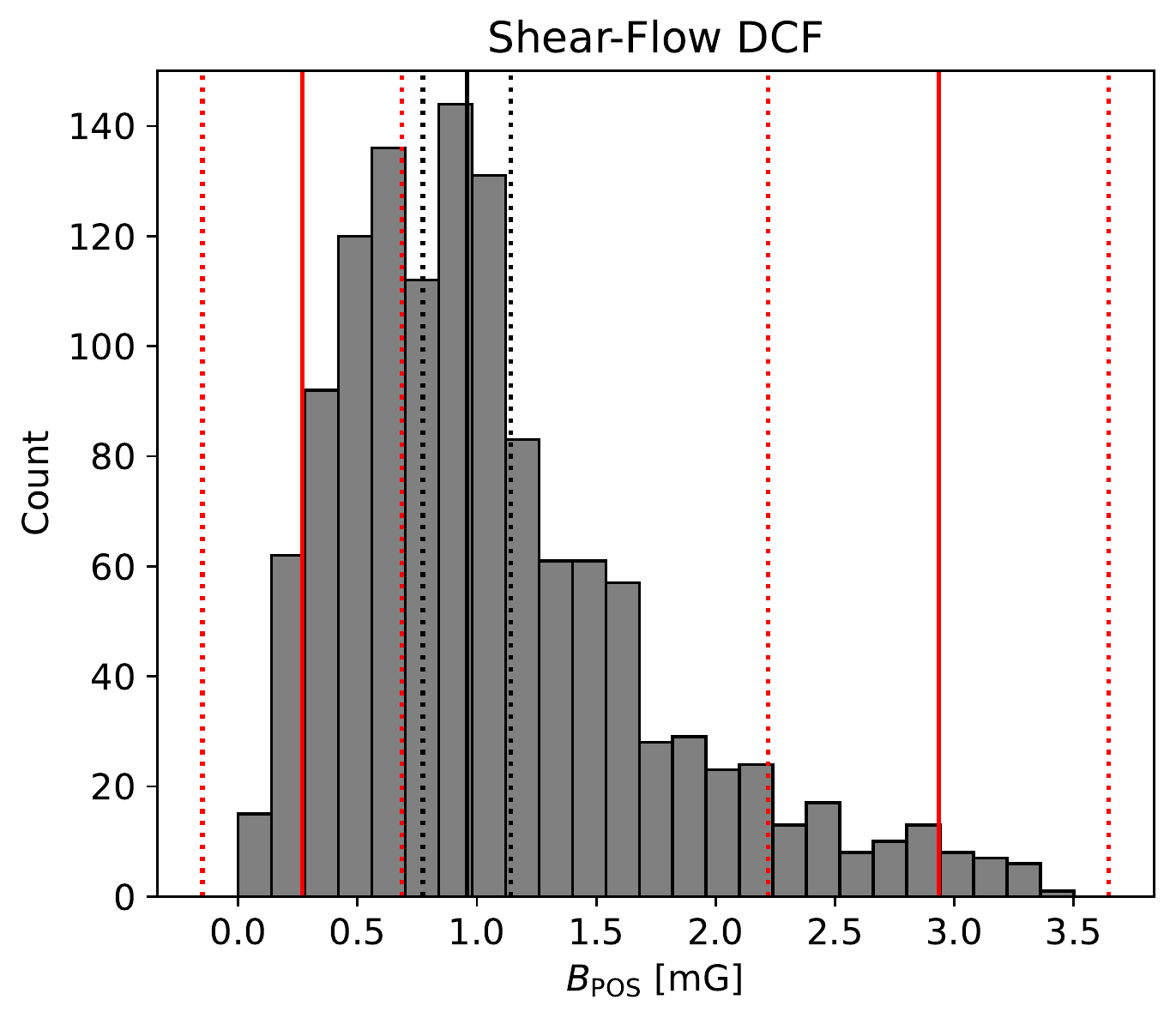}
    \caption{Distribution of POS magnetic field strength ($B_{\rm POS}$) values for the entire CND using the three DCF approximations: classical({\it Left}), large-scale flow modification({\it Middle}), shear-flow modification ({\it Right}). In all three panels, black solid vertical line indicates the median of the distribution. Red solid vertical lines to the left and right of the median indicates the 5th and 95th percentiles, correspondingly. Dotted vertical lines indicate the spread such values when uncertainties in $B_{\rm POS}$ are considered (see text for details).}
    \label{fig:Bpos_hist}
\end{figure}

Values of some variables involved in the different DCF calculations vary largely over the studied region, yielding large variations in the B-field strengths. Thus, calculating a single value (i.e., using median values for each quantity) might not be best approach. Instead, we calculate a map of B-field strength, using the maps of $\rho$, $\sigma_{v}$, $U_{0}$, $\nabla^{2}U_{0}$ and a constant-value map for $\sigma_{\phi}$. This, of course, produces a map of $B_{\rm POS}$, however because of average nature of the $\sigma_{\phi}$ value, the spatial distribution is not fully accurate. Therefore, the median values of the resulting $B_{\rm POS}$ maps are more representative. Figure \ref{fig:Bpos_hist} shows an example of the resulting distributions from which median values can be obtained, each calculated using a different DCF expression: classical ({\it Left}), large-scale flow ({\it Middle}), and shear-flow ({\it Right}). These calculations correspond to entire CND region (Section \ref{sec:CND}). B-field uncertainties are computed through the propagation of each physical variable's uncertainties using the python package \texttt{Uncertainties}\citep{LebigotUnc}. The median value (solid black line) and the range between the 5th and 95th percentiles (solid red vertical lines) are displayed. In order to account for the uncertainties in individual $B_{\rm POS}$ values, we apply the following bootstrapping-like process. First, every $B_{\rm POS}$ value is replaced with a randomly-selected value in the range [$B_{\rm POS}-\sigma(B_{\rm POS})$, $B_{\rm POS}+\sigma(B_{\rm POS})$]. Then, this process is repeated 1000 times, thus creating an ensemble of distributions of $B_{\rm POS}$ values. From this ensemble, we report values for the median, 5th, and 95th percentiles of $B_{\rm POS}$ (solid black lines in Figure \ref{fig:Bpos_hist}) and their spread as the standard deviation for the same values (dotted red lines).

\bibliography{sample631}{}
\bibliographystyle{aasjournal}



\end{document}